\documentclass[fleqn,usenatbib]{mnras}

\usepackage{newtxtext,newtxmath}

\usepackage[T1]{fontenc}

\DeclareRobustCommand{\VAN}[3]{#2}
\let\VANthebibliography\thebibliography
\def\thebibliography{\DeclareRobustCommand{\VAN}[3]{##3}\VANthebibliography}

\usepackage{graphicx}	
\usepackage{amsmath}	

\newcommand{\bahamas}{\textsc{Bahamas} }
\newcommand{\eagle}{\textsc{Eagle} }

\title[Baryon physics and RSDs]{Revisiting the effects of baryon physics on small-scale redshift space distortions}

\author[J. Kwan et al.]{
Juliana Kwan,$^{1,2}$\thanks{E-mail: jk945@cam.ac.uk}
Ian G. McCarthy$^{1}$
and Jaime Salcido$^{1}$
\\
$^{1}$Astrophysics Research Institute, Liverpool John Moores University, 146 Brownlow Hill, Liverpool, L3 5RF, UK\\
$^{2}$Department of Applied Mathematics and Theoretical Physics, University of Cambridge, Wilberforce Road, Cambridge, CB3 0WA, UK
}

\date{Accepted XXX. Received YYY; in original form ZZZ}

\pubyear{2023}

\begin{document}
\label{firstpage}
\pagerange{\pageref{firstpage}--\pageref{lastpage}}
\maketitle

\begin{abstract}
Redshift space distortions are an important probe of the growth of large-scale structure and for constraining cosmological parameters in general. 
As galaxy redshift surveys approach percent level precision in their observations of the two point clustering statistics, it is timely to review what effects baryons and associated processes such as feedback may have on small-scale clustering in redshift space. 
Contrary to previous studies in the literature, we show using the large-volume \bahamas\ hydrodynamic simulations that the effect of baryons can be as much as 1\% in the $k \sim 0.1\, h\,$Mpc$^{-1}$ range for the monopole and 5\% for quadrupole, and that this could rise to as much as 10\% at $k \sim 10 \,h\,$Mpc$^{-1}$ in both measurements. For the halo power spectra, this difference can be as much 3-4\% in the monopole on scales of $0.05 < k < 0.3 \, h\,$Mpc$^{-1}$ for 10$^{13}\,h^{-1}$M$_{\odot}$ haloes. 
We find that these deviations can be mitigated to the sub-percent level in the both the monopole and quadrupole up to $k\sim 0.3\,h$ Mpc$^{-1}$ if the baryon corrected halo masses are used to calculate the redshift space power spectra. 
Finally, we use the \textsc{cosmo-OWLS} simulation suite to explore the changes in the redshift space power spectra with different feedback prescriptions, finding that there is a maximum of 15-20\% difference between the redshift space monopole and quadrupole with and without baryons at $k \sim 1-2\,h\,$Mpc$^{-1}$ within these models. 
\end{abstract}



\section{Introduction}
\label{sec:intro}
Redshift space distortions (RSDs) appear as an apparent anisotropy in the clustering of galaxies along the line of sight produced by their peculiar velocities. 
This motion traces the local gravitational potential and thus observations of the strength of the distortions is a powerful probe of gravity and structure formation. 
Redshift space distortions are a key probe for many current and planned large volume galaxy surveys such as DESI~\citep{desi16}, Euclid~\citep{laureijs11} and the Nancy Grace Roman Space Telescope~\citep[WFIRST;][]{spergel13}, which aim to measure the linear growth rate of large-scale structure to an accuracy of $\approx$2\% error.  
While most measurements involving RSDs focus on the large-scale clustering of galaxies, where the maximum scales used are in the quasi-linear regime of $k \sim 0.2 \, h$ Mpc$^{-1}$ or $r \sim 25$ Mpc $h^{-1}$ ~\citep[see, for example,][]{alam17, satpathy17, gilmarin20, bautista21, demattia21}, analyses using smaller scales in the $\sim$$h^{-1}$Mpc range can potentially offer much more constraining power~\citep[see, for example,][]{reid14, wibking19, chapman21, yuan22, zhai22, lange23}.
However, it is important to characterize any possible impact that gas physics and the clustering of baryons may have on the redshift space clustering statistics to remove these effects as a possible source of contamination or systematic bias. 
Furthermore, many of these studies involving small-scale RSDs have reported measuring lower growth rates, $f$, and clustering amplitudes, $\sigma_8$ than other observations such as the Cosmic Microwave Background~\citep{planck18}.
Indeed, there is a growing volume of literature that has detected small ($\approx 2-3\sigma$) tensions from the Planck cosmology. 
In light of these controversies and in planning for future surveys, it becomes timely to examine the systematics that might affect these measurements, particularly as many of the models used to inform parameter estimation from RSD analyses are derived from or tested against gravity only N-body simulations. 
In particular, we explore how much the redshift space power spectrum is affected by the inclusion of baryonic content and its associated processes.

It is generally accepted that the contribution of baryonic physics may have a significant impact on clustering statistics especially on small scales, owing both to contraction via gas cooling and star formation on very small scales (e.g., \citealt{white04, zhan04}) and particularly the expansion due to feedback processes that expel gas from haloes (e.g., \citealt{vandaalen11,chisari19,vandaalen20}).
\citet{vandaalen11} showed that the amplitude of the (real space) matter power spectrum can be reduced by up to 28\% at $k=10\,h$ Mpc$^{-1}$ (1\% at $k=0.3\,h$ Mpc$^{-1}$) compared to a simulation without baryons. 
This figure can differ significantly from simulation to simulation depending on the details of the feedback implementation.
However, \citet{vandaalen20} and \citet{salcido23} have shown that these differences can be understood at approximately the percent level in terms of differences in the baryon mass fractions at a mass scale of $\sim10^{14}$ M$_\odot$.  Thus, calibration to the observed baryon fractions of groups/clusters offers a promising method for ensuring the feedback in simulations has a realistic impact on the clustering statistics \citep{mccarthy17}.    

While there have been multiple works in the literature that estimate the effect of baryons on the real space clustering, few of these make predictions that include RSD. 
\citet{hellwing16} showed using the \eagle simulations that in redshift space, the backreaction from baryons on the dark matter distribution were at most 5\% for $k > 5\,h^{-1}$ Mpc at $z=0$. 
However, \citet{kuruvilla20} measured the radial pairwise velocities of particles from a variety of hydrodynamical simulations and found a range of behaviours depending on the implementation of baryon physics (particularly the calibration strategy) in the simulations. 
They also found that the pairwise velocity can be affected by as much as 20\% on scales below $\sim h^{-1}$ Mpc and that the velocity bias between the gas content of a hydrodynamical and collisionless N-body simulation can be as much as 5\% at 10--20 $h^{-1}$Mpc. 

We will review these conclusions using the \bahamas suite of cosmological hydrodynamical simulations which are both larger in volume than \eagle (and so more appropriate for large-scale clustering) and feature improved calibration of the subgrid physics to reproduce the observed galaxy stellar mass function at $z=0.1$ and the gas fraction of clusters, which recent work has shown is crucially important.

We will also extend the analysis of~\cite{hellwing16} to include other hydrodynamic simulations with widely varying implementations of baryon physics and discuss the implications for velocity statistics.  
We find that~\cite{hellwing16} underpredicts the impact of baryons on the redshift space power spectrum; the \bahamas simulations show baryons can change the total matter power spectrum in redshift space by a few percent on scales as large as k $\sim$ 0.1\,$h\,$Mpc$^{-1}$. 

The present paper is structured as follows: in Section~\ref{sec:theory}, we introduce the redshift space clustering statistics and the simulations suites that will be used in this paper, while in Section~\ref{sec:matterpk}, we revisit the results of~\citet{hellwing16} in light of the \bahamas simulations using the total matter monopole and quadrupole power spectra. 
This section is broken up into several parts as we attempt to understand our findings in terms of 2-halo clustering in Section~\ref{sec:halo_pk} and the mean halo infall velocity and 1-halo term in Sections~\ref{sec:mean_vr} and~\ref{sec:xi1h} respectively. 
For completeness, we also explore the effects of simulation volume, redshift dependence and hydrodynamical modelling in Sections~\ref{sec:boxsize_test}, ~\ref{sec:z_dependence} and~\ref{sec:feedback} respectively. 
We summarize our findings in Section~\ref{sec:conclusions}. 

\section{Theory}
\label{sec:theory}
\subsection{Redshift space distortions}
The redshift space power spectrum can be related to the real space power spectrum via: 
\begin{equation}
\frac{d^3s}{d^3r} = \left(1 + \frac{u_r}{r}\right)^2\left(1+\frac{du_r}{dr}\right),
\label{eqn:jacobean}
\end{equation}
where $s$ is the redshift space coordinate, $r$ is the real space coordinate, $u_r$ is the scaled peculiar velocity along the line of sight (defined as $u_r \equiv v_r/H$, where $H$ is the Hubble scale), which is assumed to be in the $r$ direction. 
In the linear regime, this reduces to the Kaiser formula~\citep{kaiser87}, which relates the galaxy redshift space power spectrum, $P_{gg}$, to the total matter power spectrum, $P_{\delta\delta}$ as follows:
\begin{equation}
P_{gg}(k,\mu) = \left(b+f\mu^2\right)^2 P_{\delta\delta}(k,\mu)    
\label{eqn:kaiser}
\end{equation}
where $f$ is the linear growth rate, $b$ is the linear bias and $\mu$ is the cosine of the angle between the line of sight vector and $k$.
As equation~\ref{eqn:kaiser} shows, redshift space distortions introduce an additional degree of freedom into the power spectrum from anisotropic clustering that is dependent on the line of sight vector. 
From equations~\ref{eqn:jacobean} and~\ref{eqn:kaiser}, we can see that the clustering will be anisotropic in redshift space, as directions closest to the line of sight direction will receive the largest boosts from the velocities, so the full 2D power spectrum, $P^s(k, \mu)$, will contain useful cosmological information.

For convenience, the redshift space anisotropic 2D power spectrum is often decomposed into multipole moments via the Legendre polynomials as follows:
\begin{equation}
P^s_\ell \equiv \frac{2\ell+1}{2}\int^{+1}_{-1} d\mu \; P^s(k,\mu) \; L_\ell(\mu)
\label{eqn:multipole}
\end{equation}
where $L_\ell(\mu)$ are the Legendre polynomials and $L_0(\mu) = 1, L_2(\mu) = (3\mu^2-1)/2$ for the monopole and quadrupole moments.  
Measurements of redshift space distortions are usually expressed in terms of multipole moments for clarity and we will be focused on the monopole and quadrupole moments for the remainder of the paper, since these have the highest signal-to-noise ratio. 

In N-body simulations, we can mimick the effects of redshift space distortions by transforming real space coordinates to redshift space as follows: 
\begin{equation}
s = r + \frac{v_r (1+z)}{H(z)},
\label{eqn:particleshift}
\end{equation}
where $r$ and $v_r$ are the coordinate and velocity along the line of sight, respectively, $z$ is the redshift. 
Throughout this paper, we have assumed the plane parallel approximation in which length scales are assumed to be much smaller than the line of sight distance in $z$. 
We are then allowed to take three lines of sight, each one aligned with one of the simulation axes, and then average over the resultant two-point statistic in order to reduce the noise in our measurements. 

\subsection{Cosmological hydrodynamical simulations}
Our measurements of the redshift space power spectrum and correlation functions are derived from four sets of cosmological simulations described in Table~\ref{tab:cosmoOWLS}, namely the \textsc{Bahamas, Antilles, Eagle} and \textsc{cosmo-OWLS} suites, which include the effects of baryonic physics in the form of SPH particles and subgrid modelling.
Each of these suites contain a corresponding collisionless (`dark matter only') simulation with the same cosmology and seeded by the same initial phases in the initial conditions to facilitate the comparison. 
Additionally, the \textsc{Bahamas} simulation also has three realizations with the fiducial cosmology (WMAP9) and AGN feedback ($\log_{10}\Delta T = $ 7.8) and we take advantage of this fact whenever possible. 

\subsubsection{\textsc{Bahamas}}
The \textsc{Bahamas} (BAryons and HAlos of MAssive Systems) simulations~\citep{mccarthy17,mccarthy18} 
are a set of hydrodynamic N-body simulations run using the Smoothed Particle Hydrodynamics (SPH) code \texttt{GADGET3} SPH \citep{springel05}. 
Each box has a length of 400 $h^{-1}$Mpc per side and contains $1024^3$ dark matter particles and (initially) $1024^3$ SPH particles to represent baryons. 
The \textsc{Bahamas} simulations can be grouped into roughly two sets: those based on WMAP 9-Yr best fit cosmology~\citep{wmap9} and those with the Planck 2013 best fit cosmology~\citep{planck13}. 
In this work, we only use the simulations in the WMAP9 cosmology, with the following parameters: 
$\Omega_m= 0.2793$, $\Omega_\Lambda = 0.7207$, $\Omega_b = 0.0463$ , $h = 0.70$, $\sigma_8 = 0.8211$ and $n_s = 0.972$. 
The feedback models have been carefully calibrated to reproduce the local galaxy stellar mass function (using data from~\citealt{bernardi13,baldry12,li09}) and the gas mass fractions of galaxy groups and clusters as measured with X-ray observations \citep{vikhlinin06, maughan08, sun09, pratt09, lin12, lovisari15}. 
The~\textsc{Bahamas} simulations have been featured in a number of other works that explore the impact of baryons (and massive neutrinos) on large-scale clustering statistics such as the suppression of the real space, matter power spectrum~\citep{vandaalen20,acuto21,salcido23}, halo bias and correlation functions~\citep{pfeifer20, stafford20}  and the bispectrum~\citep{foreman20, yankelevich22}. 

We will now only briefly summarise the baryon implementation in the \textsc{Bahamas} project; we refer the interested reader to~\citet{mccarthy17} for a fuller discussion. 
All simulations include photoionisation calculated by the \textsc{Cloudy} package~\citep{ferland98} that accounts for CMB and the UV/X-ray background~\citep{haardt01}. 
The subgrid prescription for star formation and stellar feedback follows~\citet{schaye08} and~\citet{dallavecchia08} respectively, while the cooling scheme is implemented according to~\citet{wiersma09}. 
AGN feedback is achieved through accretion of gas onto a supermassive black hole via the scheme described in~\citet{booth09}; the accreted energy is parcelled out into chunks of $\Delta$T, a parameter that describes the temperature increase in the surrounding SPH particles. 
Higher values of $\Delta$T implies more energetic (but less frequent) feedback events when the same amount of gas is being expelled. 
Since $\Delta$T (and thereby AGN feedback in general) was shown to be one of the key parameters governing the cluster gas mass fraction and hence important observables such as the X-ray luminosity and thermal SZ flux~\citep{lebrun14}, this is explored in three scenarios in \textsc{Bahamas}, named low AGN, fiducial and hi AGN, which corresponds to $\log_{10}\Delta$T = 7.6, 7.8 and 8.0 respectively.

\subsubsection{\textsc{Antilles}}
The \textsc{Antilles} suite of simulations, introduced in~\citet{salcido23}, is a large set cosmological hydrodynamical simulations spanning a very wide range of the baryonic feedback models in sufficiently large volumes designed for large scale structure studies. The \textsc{Antilles} suite adopts the same flat $\Lambda$CDM cosmology as \textsc{Bahamas}, consistent with the WMAP 9-year results~\citep{wmap9}, as well as the same mass particle resolution, corresponding to (initial) masses of $m_\mathrm{g} = 1.09 \times 10^9 \mathrm{M}_\odot$ and $m_\mathrm{dm}=5.51 \times 10^9 \mathrm{M}_\odot$ respectively.
An important improvement in \textsc{Antilles}, is the use of the state-of-the-art `Anarchy' SPH formulation used in \textsc{Eagle}. The improvements within `Anarchy' include the use of the pressure-entropy formulation of SPH derived by \cite{hopkins13}, the artificial viscosity switch from \cite{cullen10}, an artificial conduction switch similar to that of \cite{price08}, the $\mathcal{C}_2$ kernel of \cite{wendland95}, and the time-step limiters of \cite{durier12}. 

\textsc{Antilles} employs the same subgrid-physics prescriptions for radiative cooling and photoheating, star formation, time-dependent stellar mass loss due to winds from massive stars and AGB stars, core collapse supernovae and type Ia supernovae, kinetic wind stellar feedback model, and thermal feedback from AGN as the the \textsc{Bahamas} simulations. 

Here we present three calibrated \textsc{Antilles} simulation boxes that used Gaussian process emulators trained on 200 `Anarchy' SPH \textsc{Antilles} feedback variations in a box size of 100 Mpc/$h$, to model how the galaxy stellar mass function and cluster gas fractions change as a function of the subgrid parameters. We use these emulators to the calibrate subgrid-physics parameters to match the observed galaxy stellar mass function and the observed gas fraction in groups and clusters.

The calibrated parameters were used to produce three simulation boxes using the same mass resolution, but in increasing box sizes of 100 Mpc/$h$, 200 Mpc/$h$ and 800 Mpc/$h$. The details of these calibrated \textsc{Antilles} simulations are listed in Table~\ref{tab:cosmoOWLS}. 
These simulations were used to investigate the effects of box size on the redshift space clustering (detailed in Section~\ref{sec:boxsize_test}) since the volumes neatly bracket the \bahamas simulations.

\begin{table*}
    \centering
    \caption{Set of simulations used in this study. Each simulation also has a dark matter only counterpart with which we use to make comparisons. Note that the values of $\Delta$T are not directly comparable between simulation suites because of differences in the feedback implementation (as described in the text). Nonetheless, we include it here as it is the main driver for the cluster baryon fraction and hence changes to the (real space) matter power spectrum. 
    }
    \label{tab:cosmoOWLS}
    \begin{tabular}{c|c|c|c|c|c|c|c}
        \hline
        Name & L [$h^{-1}$Mpc] & N & Cosmology & Cooling & Star formation & SN feedback & $\log_{10}\Delta$T \\
        \hline
        \bahamas fiducial & 400 & $2 \times $1024$^3$ & WMAP9 & Yes & Yes & Yes & 7.8 \\
        \bahamas low AGN  & 400 & $2 \times $1024$^3$&  WMAP9 & Yes & Yes & Yes & 7.6 \\
        \bahamas hi AGN   & 400 & $2 \times $1024$^3$&  WMAP9 & Yes & Yes & Yes & 8.0 \\
         \eagle REF        & 67.77 & $2 \times $1504$^3$&  Planck13 & Yes & Yes & Yes & 8.0 \\
         \textsc{cosmo-OWLs} NOCOOL    & 400 &$2 \times $1024$^3$& WMAP7 &No  & No  & No   & N/A  \\
         \textsc{cosmo-OWLs} REF       & 400 &$2 \times $1024$^3$& WMAP7 &Yes & Yes & Yes  & N/A \\
         \textsc{cosmo-OWLs} AGN       & 400 &$2 \times $1024$^3$& WMAP7 &Yes & Yes & Yes  & 8.0 \\
         \textsc{cosmo-OWLs} AGN 8.5 & 400 &$2 \times $1024$^3$& WMAP7 &Yes & Yes & Yes  & 8.5 \\
         \textsc{cosmo-OWLs} AGN 8.7 & 400 &$2 \times $1024$^3$& WMAP7 &Yes & Yes & Yes  & 8.7 \\
         \textsc{Antilles}-L100              & 100 &$2 \times $256$^3$& WMAP9 &Yes & Yes & Yes & 8.01 \\ 
         \textsc{Antilles}-L200              & 200 &$2 \times $512$^3$& WMAP9 &Yes & Yes & Yes & 8.01 \\
         \textsc{Antilles}-L800              & 800 &$2 \times $2048$^3$& WMAP9 &Yes & Yes & Yes & 8.01 \\
         \hline
    \end{tabular}
\end{table*}

\subsubsection{\textsc{Eagle}}
The \textsc{Eagle} (Evolution and Assembly of GaLaxies and their Environments) suite of hydrodynamical simulations~\citep{schaye15, crain15} aim to capture the formation and evolution of galaxies and supermassive black holes by tuning the subgrid physics parameters to reproduce the observed galaxy stellar mass function at $z=0.1$ and the size-stellar mass relation of local galaxies. 
These simulations consist of a number of different volumes at different resolutions but we will only focus on the largest of these (\textsc{Eagle} REF) with a box size of $L=67.77$ Mpc/$h$, and 1504$^3$ dark matter particles and an equal number of baryons for the purposes of comparison. The cosmological parameters of the \textsc{Eagle} simulations are fixed at the Planck 13 best fitting parameters~\citep{planck13}: $\Omega_m= 0.307$, $\Omega_\Lambda = 0.693$, $\Omega_b = 0.04825$ , $h = 0.6777$, $\sigma_8 = 0.8288$ and $n_s = 0.9611$ for all variations. 
The \textsc{Eagle} suite used the `Anarchy' SPH formulation. 
In terms of the gravity solver and the subgrid modelling, \eagle is quite similar to \textsc{Bahamas},  \textsc{Antilles} and \textsc{cosmo-OWLs} in that it also uses \textsc{Gadget3}. 
However, some notable changes include - the energy feedback from star formation is modelled as thermal rather than kinetic, the Bondi-Hoyle accretion rate is not adjusted by an additional boost factor, $\alpha$, and there are metallicity dependent thresholds for star formation.

\subsubsection{\textsc{cosmo-OWLS}}
As an extension of the \textsc{OWLS} (Overwhelmingly Large Simulations) project~\citep{schaye10}, the aim of \textsc{cosmo-OWLs}~\citep{lebrun14, mccarthy14} is to provide hydrodynamic simulations in a cosmological volume and explore the impact of different aspects of the subgrid physics, such as cooling and AGN feedback on cluster observables. 
The resultant wide ranging behaviour found under the various \textsc{cosmo-OWLS} subgrid variations directly lead to the calibration strategy of the \textsc{Bahamas} suite; in order to reproduce large scale structure observables, it would be necessary to match the measured cluster gas fractions and galaxy stellar mass function via calibration of the subgrid parameters. 
We use the version of these simulations with a WMAP-7 like cosmology, namely $\Omega_m= 0.272$, $\Omega_\Lambda = 0.728$, $\Omega_b = 0.0455$, $h = 0.704$, $\sigma_8 = 0.81$ and $n_s = 0.967$. 
The size and resolution of \textsc{cosmo-OWLS} is the same as the \textsc{Bahamas} suite, with 400 $h^{-1}$Mpc per side and 1024$^3$ dark matter particles and 1024$^3$ SPH particles. 
In order to isolate the impact of various baryonic prescriptions, there are five `flavours' in the \textsc{cosmo-OWLS} suite, as described in Table 1, in which various gas physics, such as supernova (SN) and AGN feedback, cooling etc., have been toggled on and off in each run. 
The implementation of each effect is the same as those in the \textsc{Bahamas} simulations, which we have described in an earlier section.

\begin{figure*}
\includegraphics[width=0.8\textwidth]{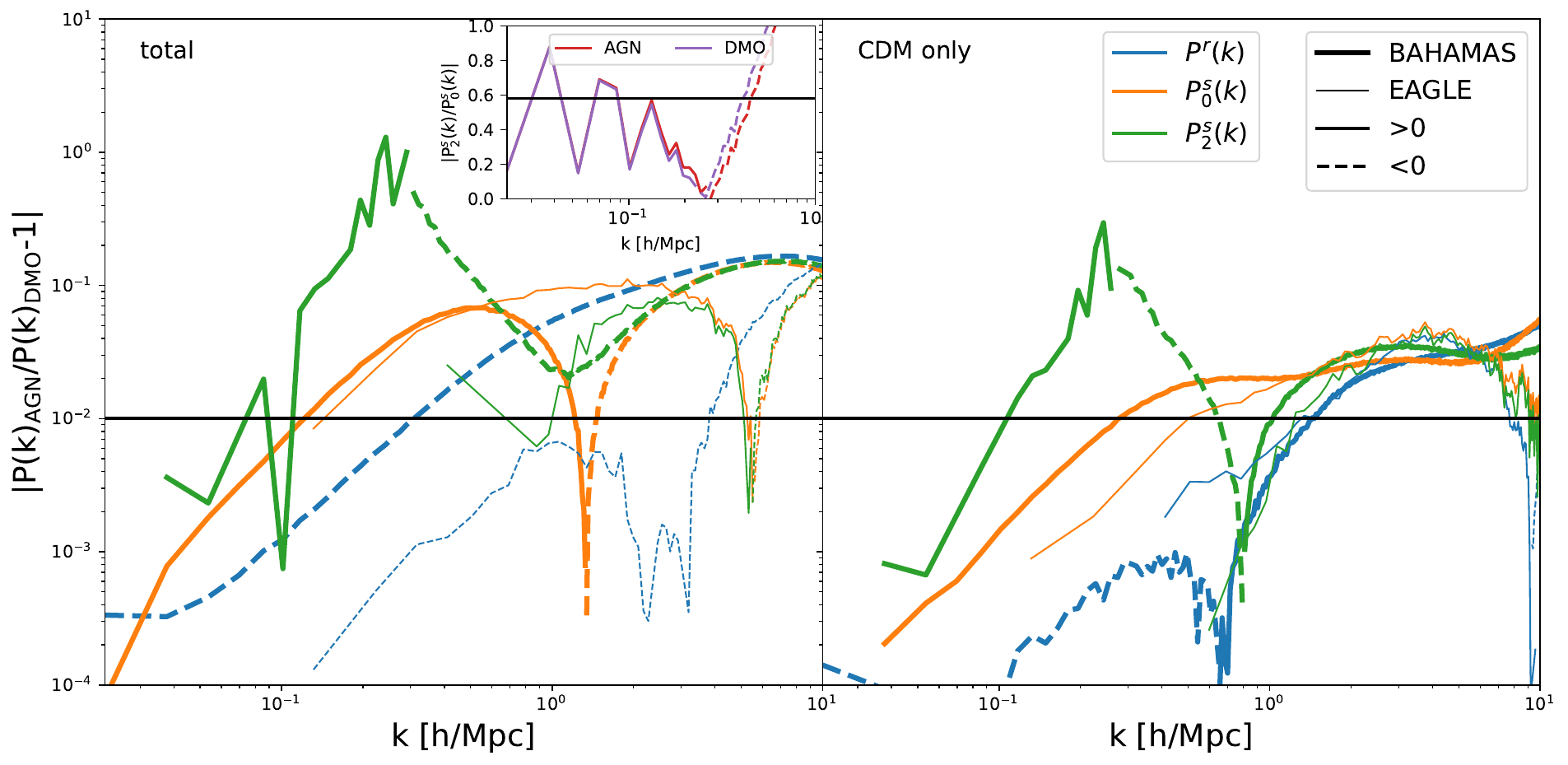}
\caption{The effect of baryons on the total (left) and dark matter (right) real and redshift matter power spectra of the \textsc{Bahamas} and \textsc{Eagle} simulations at $z=0$. Each measurement is expressed as a ratio to the corresponding dark matter only multipole. Solid lines denote positive values of the ratio, while dashed lines represent negative values. Insert: the ratio of the quadrupole to monopole moment in \bahamas measured from both the fiducial AGN (red) and dark matter only (purple) simulations. The black line shows the prediction from the Kaiser formula. Again negative values are indicated by the dotted lines. We can see that the redshift space monopole and quadrupole moments in the \bahamas simulations (and also the real space clustering) show an enhanced baryonic effect relative to \eagle.}
\label{fig:eagle_comparison}
\end{figure*}

\begin{figure}
\centering
\includegraphics[width=0.5\textwidth]{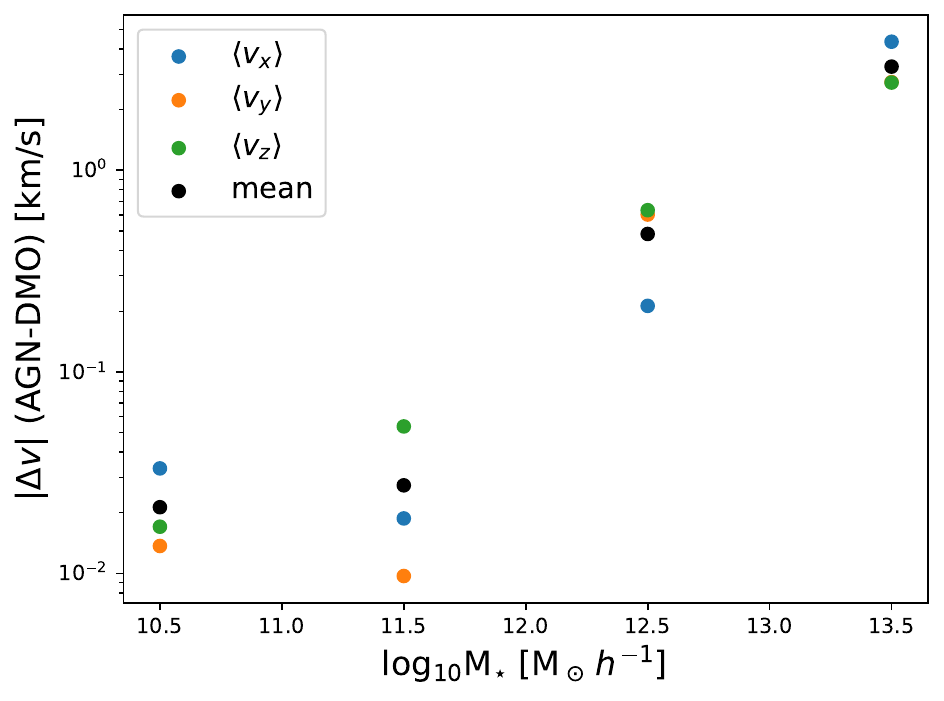}
\caption{The difference in the mean peculiar velocities of haloes identified in the \bahamas dark matter only  and the \bahamas AGN tuned simulations in bins of stellar mass. Stellar masses have been assigned to the dark matter only haloes using a process of bijectively matching the IDs of each halo member particle with its equivalent in the AGN tuned simulation (as explained in the text). }
\label{fig:halo_velocity_difference}
\end{figure}

\section{Baryonic effects on the redshift space matter power spectrum}~\label{sec:matterpk}
In Figure~\ref{fig:eagle_comparison}, we show the total and dark matter redshift space monopole and quadrupole moments measured from the \textsc{Bahamas} fiducial simulation and for comparison, the results from the \textsc{Eagle} REF simulation as used by~\citet{hellwing16}. 
We find several important deviations from their results: that the total effect of baryons on the monopole matter power spectrum is $\sim$10\% up to $k \approx 10\;h$ Mpc$^{-1}$, that the effect extends to much larger scales, particularly for the quadrupole moment and that the bulk mean halo peculiar velocity, $\Delta v$, exceeds their reported value of $\sim 1$ km/s for the most massive haloes in the \bahamas simulations.  
Note that here we have quoted the numbers from the total monopole moment (left panel) whereas~\citet{hellwing16} presents the dark matter component (or backreaction, right panel) of the total matter power spectrum, which generally reduces the size of the effect. Nonetheless, the backreaction is still more significant on large scales in the \bahamas simulations than what is seen in the \eagle simulations in their paper.
Just as the ratio between simulations is declining in the \textsc{Eagle} scenario, the same ratio taken in the \textsc{Bahamas} simulations peaks at $k \approx 0.2 \; h$ Mpc$^{-1}$.  
Note that the effect on the quadrupole appears to be particularly strong, but this is due to a change in sign in the transition from linear, Kaiser model smearing to the non-linear `Fingers-of-God' effect (i.e., the dominator in the ratio passes through zero, leading to a spike at this scale). 
We have included an insert in Figure~\ref{fig:eagle_comparison} in which we show the ratio of the \bahamas quadrupole to monopole moments with and without baryons to properly portray the size of the effect and demonstrate that it is indeed caused when the quadrupole moment goes through zero. We have also included the Kaiser limit from Equation~\ref{eqn:kaiser} for reference.
This spike in the quadrupole ratio is seen in the \bahamas and not the \eagle simulation because the smaller box size of the latter suppresses features of the power spectrum on large scales. 
Furthermore, on much smaller scales, $k > 2-3\,h\,$Mpc$^{-1}$, we start to see the effects of AGN and stellar feedback on the halo profiles in redshift space. 
Haloes in the dark matter only versions of these simulations are more concentrated because AGN feedback and, to a lesser extent, stellar feedback move material within the halo and act to decrease its concentration.

Figure~\ref{fig:halo_velocity_difference} shows the difference in the mean bulk halo velocities measured from the \bahamas simulations with and without the effect of baryons. 
To obtain the stellar masses for the dark matter only haloes, we use the same catalogue of matched haloes as in~\citet{stafford20}, which was produced by bijectively matching the \textsc{subfind} halo catalogues between the fiducial AGN and DM only \bahamas simulations at $z=0$.  
Since these simulations form a pair with the same initial phases, we can associate most haloes from the fiducial AGN run with their counterpart from the dark matter only run by cross matching the IDs of dark matter particles within both sets of haloes. 
We require that 50\% or more of the particles to be present in both haloes across the two simulations before deciding that they are a match. 
We then aggregate each of these haloes into bins of 1 dex in stellar mass, starting from 10$^{10}$ M$_\star$ $h^{-1}$, which is resolved by $\sim$10 stellar particles, calculate their mean peculiar velocities and subtract them for each direction. 
We found that while the lower stellar mass bins were consistent with findings of~\citet{hellwing16}, namely that the absolute difference in the bulk halo velocities with and without the effect of baryons is $< 1$km/s, for the haloes with $> 10^{13}$ M$_{\star}\, h^{-1}$,  this rises to $> 4$km/s.

There are several ways in which the \eagle simulations differs from the \bahamas suite, but most significantly \eagle has weaker AGN feedback which results in a higher gas fraction in clusters with masses $>10^{14}\;h^{-1}$. 
This results in an overall reduced baryon suppression of the real space matter matter power spectrum~\citep{hellwing16, vandaalen20} compared to \bahamas.
Figure~\ref{fig:eagle_comparison} shows that this underestimate of the baryon suppression is also passed onto the redshift space two-point statistics as well.  
Additionally, the cosmology and hence the growth rate varies between these simulations as well; the \bahamas suite adopts WMAP9 parameters while \eagle uses a Planck13 best fit cosmology. 
This gives a slightly lower growth rate of $\sim$0.49 versus $\sim$0.52 respectively.
When propagated to the monopole and quadrupole power spectra via the Kaiser formula, this implies that to first order the \eagle monopole and quadrupole power spectra should be 2\% and 6\% stronger relative to \bahamas.
We explore the cause for the difference in the redshift space clustering between the hydro and dark matter only simulations in the following sections, as we examine the impact of feedback on the clustering of haloes both between themselves in the halo-halo power spectrum and internally with the 1-halo central-satellite correlation function.

\begin{figure*}
    \centering
    \includegraphics[width=0.8\linewidth]{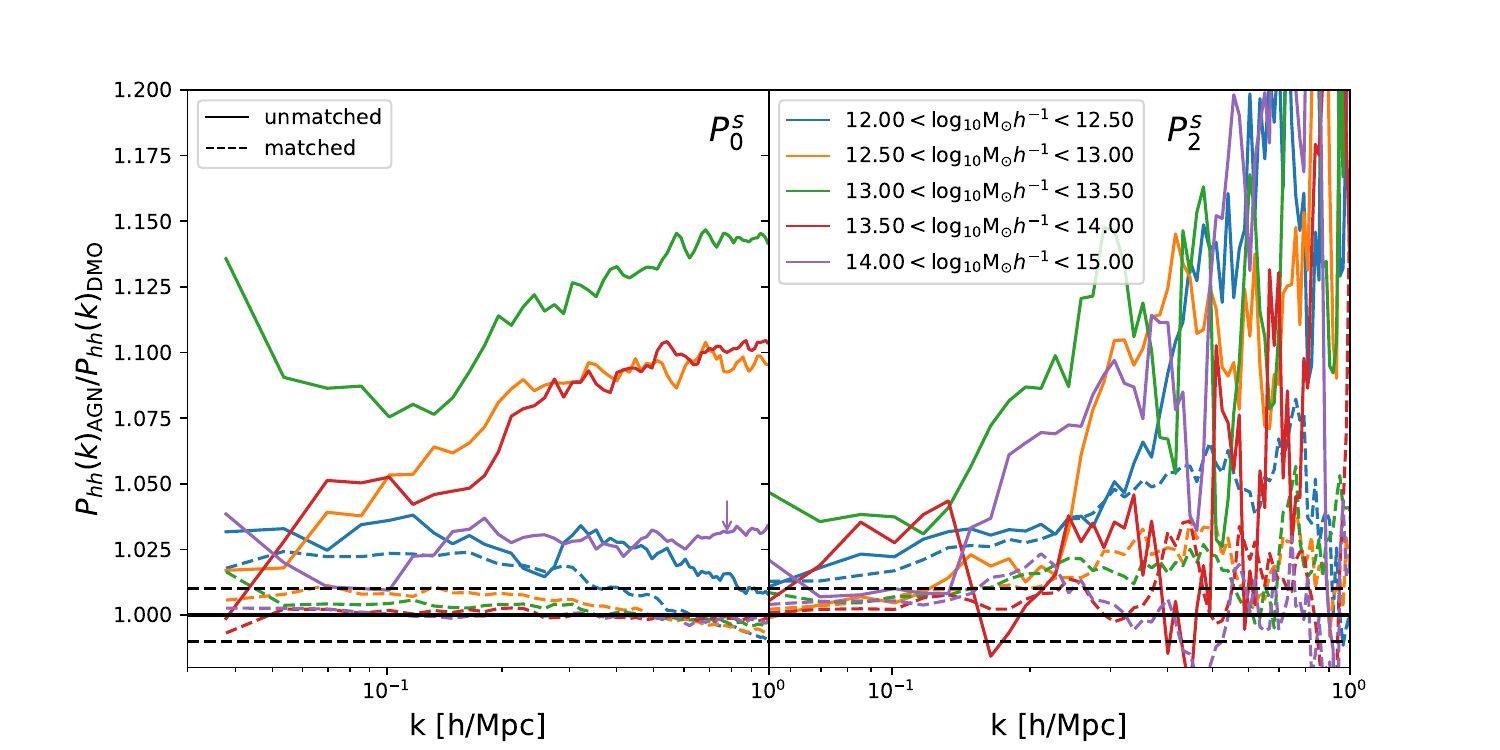}
    \caption{Baryonic effect on the halo-halo monopole (left) and the quadrupole (right) measured from the \bahamas simulations. The solid curves are calculated using mass cuts applied individually to each simulation while dashed lines use a matched sample of haloes; the selection is now performed on the halo masses measured in the \bahamas fiducial AGN simulation and then we locate their equivalents in the \bahamas DM only simulation. The halo mass binning for each curve is shown in the legend and 1\% deviations have been marked in dotted lines. Note that the monopole ratios have been smoothed with a moving 3-point average (5-point for the quadrupole) in order to bring out the large scale features. Arrows indicate the $k$ scale at which the shot noise becomes dominant in the \bahamas AGN monopole moment.}
    \label{fig:halo_mass_selected}
\end{figure*}

\begin{figure*}
    \centering
    \includegraphics[width=0.8\linewidth]{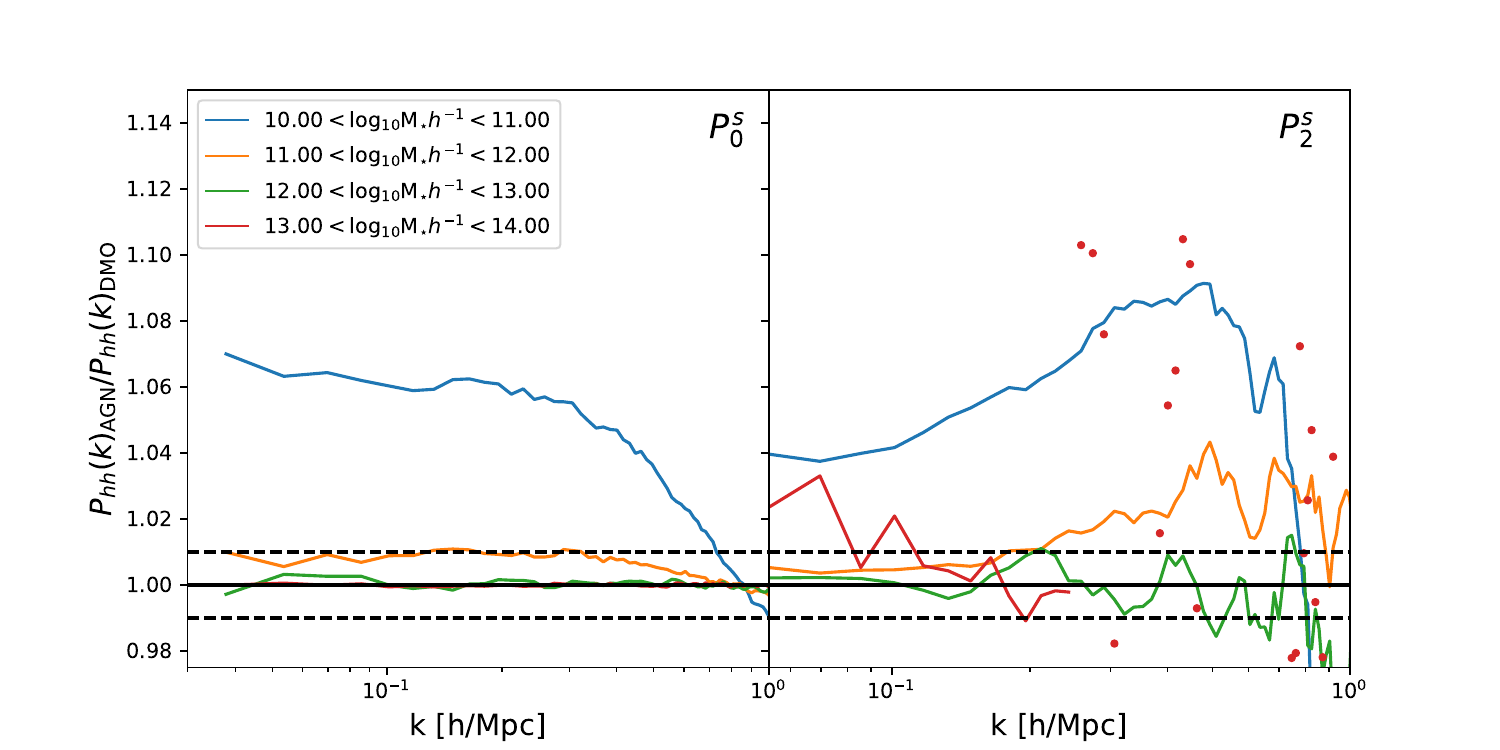}
    \caption{Difference in the ratio between the monopole (left) and quadrupole (right) with and without the effect of baryons for a stellar mass selected set of haloes; the matching procedure between the \bahamas dark matter only simulation and the \bahamas fiducial AGN simulation allows us to supply the former with the appropriate stellar masses as well. To avoid crowding the plot, we have used dots to represent the quadrupole in the highest stellar mass bin where the measurements are noisy.}
    \label{fig:stellar_mass_selected}
\end{figure*}

\subsection{Effect on large scale clustering: Halo mass dependence}~\label{sec:halo_pk}
In order to understand the change in large-scale clustering between the gravity only and hydrodynamic simulations shown in the previous section, we now investigate the impact of baryons on the halo-halo monopole and quadrupole moments measured in redshift space. 
We measure the power spectrum between haloes divided into five mass bins from halo catalogues built from the \bahamas simulations using \textsc{Subfind}~\citep{springel05, dolag09}.
\textsc{Subfind} identifies haloes by running a Friends-of-Friends (FOF) algorithm with a link length of $b=0.2$ to first identify groups and then applies a spherical overdensity finder with an unbinding procedure to remove spuriously linked particles with total energies that exceed the binding energy of the halo. 
We impose a further constraint by choosing to use only FOF groups in our comparison; this is important for our discussion in the next section involving the infall velocities of central haloes and their constituent particles.
We have chosen to divide the haloes using their FOF group masses into bins of 0.5 dex in $h^{-1}$M$_{\odot}$ starting from 10$^{12}\,h^{-1}$ M$_{\odot}$ until the final bin at 10$^{14}\,h^{-1}$ M$_{\odot}$, in which we take haloes up to 10$^{15}\,h^{-1}$ M$_{\odot}$. 
We found that this sampling was a good compromise between reducing scatter from Poissonian shot noise and being able to clearly identify features between each bin. 
For comparison, we also calculate an equivalent halo catalogue in the \bahamas DM (dark matter) only simulation.

Figures~\ref{fig:halo_mass_selected} and~\ref{fig:stellar_mass_selected} show the changes to the monopole and quadrupole moments in redshift space as a result of the inclusion of baryonic physics. 
For Figure~\ref{fig:halo_mass_selected}, the mass binning is performed on the total halo mass for each simulation with the halo mass defined as the sum of all the particles belonging to the halo, which in the case of the fiducial AGN simulation includes the gas and stars, as identified by \textsc{Subfind}. 
Note that because of limited sample numbers, the ratio cannot be interpreted over the entire $k$ range shown for $10^{14} < h^{-1}$ M$_{\odot} < 10^{15}$ mass haloes. To guide the reader, we have also marked the $k$ scale at which the shot noise contribution to the monopole moment of the \bahamas AGN fiducial model becomes dominant. The shot noise has been calculated as $1/\bar{n}$, where $\bar{n}$ is the number density of haloes in that mass bin.
We can see that there is a substantial difference between the halo-halo power spectra across simulations, even on the very largest scales.  
This is greatest for haloes in the range 13.0 $< \log_{10}$ M$_{\odot}h^{-1} < $ 13.5 for which the difference is 7-10\% at k$<0.1 h^{-1}$Mpc in the monopole (3\% in the quadrupole) and rises to $\sim$13\% (15\%) on smaller scales. 
Some of this difference is due to the sample selection, in that the inclusion of baryon has changed the halo abundance so the same haloes are not being compared across the simulations; in fact this very mass range in which we see the greatest difference between the power spectra is also the same range at which the peak difference in the halo mass functions occurs as first noted by~\citet{vandaalen14} (see figure 14 of~\citealt{pfeifer20} for the specific \bahamas mass function). 
This is because haloes in the 13.0 $< \log_{10}$ M$_{\odot}h^{-1} < $ 13.5 range in the fiducial AGN simulation are the most susceptible to AGN ejection. 
\citet{schaller15} showed that for the \textsc{Eagle} simulations, the effect of gas physics is to monotonically lower the halo masses except for the very largest clusters at $>10^{14}$ $h^{-1}$ M$_\odot$, meaning that these objects are now rarer and so the halo bias is increased. 
For low mass haloes of $\sim 10^{12}$ M$_{\odot}h^{-1}$, the baryons and even the dark matter particles inside the halo are disrupted by feedback processes and the halo radius shrinks, however, as the halo mass increases, stellar feedback becomes less effective and the ratio of halo masses, M$_{\rm AGN}$/M$_{\rm DMO}$, approaches the universal dark matter fraction. 
At even higher masses ($\sim 10^{14}$ M$_{\odot}h^{-1}$), the mass ratio approaches unity as the large gravitational potential of the cluster prevents baryon expulsion from AGN feedback. 
We see a similar effect in \bahamas, except the mass ratio, M$_{\rm AGN}$/M$_{\rm DMO}$, does not continue to fall for lower masses, rather there is an upturn at $\sim 10^{12.8}$ $h^{-1}$ M$_\odot$ and these haloes now have approximately the same halo mass with or without gas physics. 
This is because the stellar feedback in \textsc{Eagle} is stronger than in \bahamas, but \eagle has weaker AGN feedback.  
This relationship between M$_{\rm AGN}$ and M$_{\rm DMO}$ is reflected in the monopole moment in Figure~\ref{fig:halo_mass_selected}, in which both the lowest and highest mass bins exhibit the smallest change in behaviour in the redshift space monopole since they experience the smallest change in mass.

Suppose the effect of the halo mass function were to be removed - how much of this difference would remain? 
Figures~\ref{fig:halo_mass_selected} and~\ref{fig:stellar_mass_selected} also show the ratio between the multipole power spectra obtained from fiducial AGN set of halo catalogues to the multipole power spectra as measured from their equivalent haloes in a dark matter only simulation for the same total halo mass and stellar mass cuts respectively. 
We use the same matched halo catalogues as in Figure~\ref{fig:halo_velocity_difference} for the comparison between the mean bulk halo velocities. 
The mass selection is first applied to the fiducial AGN simulation and then these haloes are mapped to their equivalents in the dark matter only simulation which may have any mass (down to 50\% of the matched halo). 
A small number of haloes will not have a direct analogue of course, especially those with a lower mass, which may have been disrupted or undergone a merger. 
For consistency, the halo sample from the dark matter only simulation remains the same regardless of whether we are comparing to matched or unmatched haloes from the \bahamas AGN fiducial simulation.
Matching the halo catalogues in this way allows us to minimize the difference between the \bahamas fiducial AGN and gravity only runs, since we have fully accounted for changes in the halo mass.

Figure~\ref{fig:halo_mass_selected} demonstrates that there is a only a percent level difference in the large scale halo bias of both monopole and quadrupole moments if we used matched catalogues to compare the fiducial AGN \bahamas simulation with its dark matter only equivalent. 
For the lowest mass range, however, a difference of 2.5\% remains in the large scale halo bias even for the matched case. We have determined that this is due to 
missing haloes in the AGN fiducial case causing the matched halo mass function to be systematically lower than the unmatched version by about 2-4\% in this mass range, because some of the smallest haloes do not have a dark matter only counterpart. Some of these cases would be due to limitations in the mass resolution in the \bahamas simulations, while others could have undergone a physical process that destroys the halo.
Again, while there is no discernible trend with increasing halo mass in the matched quadrupole power spectra  (Figure~\ref{fig:halo_mass_selected}, right), there seems to be a steady enhancement of the difference in the quadrupole moments over $0.3 < k < 1 \, h$ Mpc$^{-1}$ up to $\sim$15\%. 
This implies that a simple remapping of the halo mass function to its equivalent under gas physics is sufficient to account for large scale changes ($k < 0.3\,h^{-1}\,$Mpc) when baryons are included if we have access to their equivalent haloes that would have evolved under gravity only conditions. 
We have also checked that this remains true when using a matched sample from the other simulations in the \bahamas suite.
Our results are consistent with the trend seen in the real space halo power spectrum~\citep{yankelevich22}. 
Furthermore,~\citet{yankelevich22} show that the (real space) halo bispectrum is more sensitive to these changes in the halo mass, where the halo profile makes an appreciable impact. 
It would be interesting to see if the same holds true for the redshift space bispectrum, but we leave this for a future exploration. 

In Figure~\ref{fig:stellar_mass_selected}, we have shown the difference in halo bias calculated for a stellar mass selected sample of haloes. 
As in the previous section, we use the matched catalogues between the AGN tuned and dark matter only simulations. Note that the shot noise contribution is only significant on $k > 1\,h^{-1}Mpc$, which is beyond the relevent scales for this analysis. 
Except for the very lowest stellar mass bin, none of the ratios of the monopole power spectra show a significant difference between two simulations at greater than the percent level. 
The same holds for the quadrupole, however, in addition to the 6\% difference in the lowest stellar mass bin, there is also a 2\% offset for the greatest stellar mass haloes as well. 
For the monopole, this is about two times larger than the effect seen in the halo mass selected matched sample, which suggests that not all of the effect is accounted for by the limitation in mass resolution of the simulations. 
Furthermore, the median halo masses for the lowest stellar mass bin are 10$^{11.64}\,h^{-1}M_{\odot}$ and 10$^{11.67}\,h^{-1}M_{\odot}$ for the dark matter only and AGN tuned models respectively, which corresponds to $\sim$80 particles so these should correspond to resolved haloes. 
Still Figure~\ref{fig:stellar_mass_selected} suggests that the monopole and quadrupole moments of a stellar selected sample will not show more than 3\% deviation between a dark matter only and a full hydrodynamical simulation on scales $> 0.1 h$Mpc$^{-1}$.

\subsection{Effect on small scale clustering: Halo profiles}
Having established that the change in halo masses under baryonic physics is the main driver of the difference between the redshift space 2-halo term in the \bahamas fiducial AGN simulation and the DM only simulation on large scales, in this section, we take a closer look at the effect of baryons on the small scale clustering inside haloes via the mean infall velocity, $v_{12}(r)$ and the redshift space 1-halo correlation function, $\xi_{1h}(r)$. 

\subsubsection{Halo mean infall velocity: Halo velocity profile}\label{sec:mean_vr}
The pairwise radial velocity, $v_r$, between two objects with peculiar velocities, $v_{1, \rm pec}$ and $v_{2, \rm pec}$, is defined as:
\begin{equation}
 v_{12}(r) = v_{\rm 1, pec} \cdot \hat{r}_{12} - v_{\rm 2, pec} \cdot \hat{r}_{12}, 
\end{equation}
where $r_{12}$ is their radial separation. 
In order to study the impact of baryonic effects on the halo velocity profile, we will be considering correlations between FOF groups identified by \textsc{subfind} and the particles contained within them; this is to avoid complications involving the identification of central and satellite subhaloes and because we want the center of the halo to be at rest with respect to its member particles in order to differentiate between the bulk motion of the halo and the internal motion of the halo particles.
The halo velocity is defined as the average of the comoving peculiar velocities of its member particles; this will be comprised of the bulk motion of the halo as well as a component due to infall from the gravitational potential of the halo, while the random virial motions are expected to be averaged out.
The pairwise radial velocity is a major ingredient in understanding redshift space distortions, and since we have chosen pairs between the halo centre (defined as the potential minimum) and the particles that are bound to that halo, it should be a direct measure of infall as a function of halo radius. 
 
In this section, we consider how the mean pairwise velocity, $\left < v_{12} \right>$, around haloes might change between an N-body simulation with full hydrodynamics and one with gravity only.
We use the \texttt{Halotools} package~\citep{hearin17} to measure the mean pairwise radial velocity around groups in the \bahamas simulations.

\begin{figure*}
    \centering
    \includegraphics[width=\linewidth]{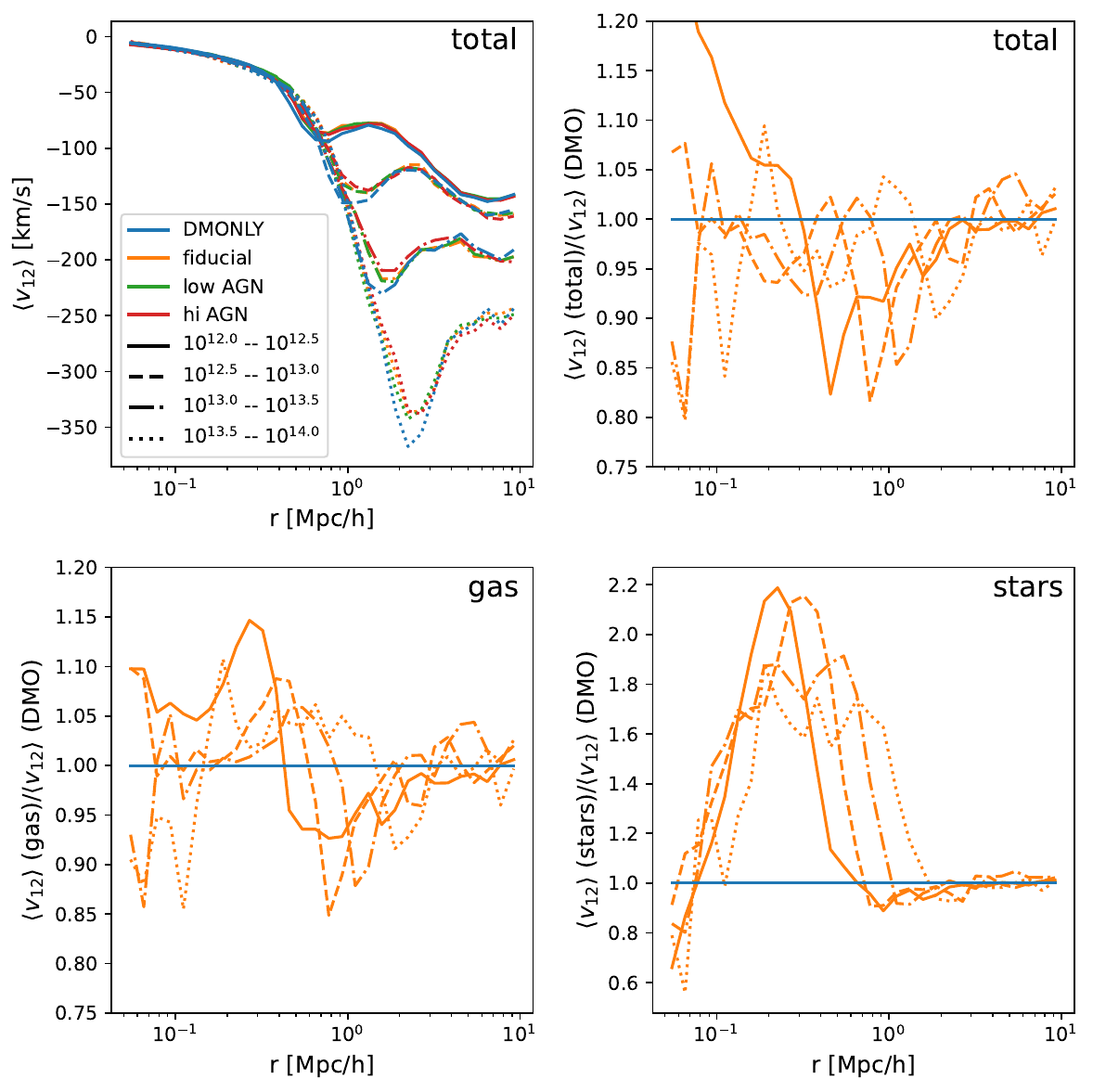}
    \caption{The mean infall velocity of halo particles with and without gas physics in the \bahamas simulations across four halo mass bins: 10$^{12} < h^{-1}$M$_{\odot} \leq $ 10$^{12.5}$ (solid), 10$^{12.5} < h^{-1}$ M$_\odot \leq $10$^{13.0}$ (dashed), 10$^{13.0}< h^{-1}$M$_{\odot} \leq $10$^{13.5}$ (dot-dashed) and 10$^{13.5} h^{-1}< $M$_{\odot} \leq $10$^{14.0}$ (dotted). The top left plot shows the total $\left < v_{12} \right >$ as a function of halo radius for all halo particles, while the others are shown as a ratio with respect to the dark matter only simulation. Only the fiducial AGN simulation has been kept for the ratio plots, as the other AGN models show very similar behaviour. Additionally, the behaviour of $\left < v_{12} \right >$ using only DM particles (i.e. the backreaction) is so qualitatively similar to the total $\left < v_{12} \right >$ that we have opted to only show the latter.}
    \label{fig:mean_vr}
\end{figure*}

In Figure~\ref{fig:mean_vr} we show $\left < v_{12} \right >$ calculated between central haloes and halo particles of the \bahamas simulations as a function of the distance to the halo center for the total matter content. 
This is done for four bins in halo mass spanning 0.5 dex (as in Figure~\ref{fig:halo_mass_selected}) for each of the \bahamas simulations listed in Table~\ref{tab:cosmoOWLS} and a dark matter only counterpart. 
The top left panel shows the total $\left < v_{12} \right >$ for all halo particle types and all \bahamas simulations, while the other three panels are ratios with respect to the dark matter only simulation for the fiducial AGN model using a different species each time, namely: all particles (total), gas particles only and star particles only. 
From the top left panel, we can immediately see that the velocity profiles of the haloes within each mass bin do not seem to differ a great deal across simulations; in fact for clarity, we have omitted the other \bahamas variations when plotting the ratios in the three other panels of Fig~\ref{fig:mean_vr}. 
Also since no unbinding procedure is performed, some interloper particles and spurious linkage causes measurements of the mean infall velocity to continue to be defined beyond the halo radius. 
We have checked that the different AGN models only induce a modest 3-4\% change relative to the fiducial model in the halo mean radial velocity at most for the gas and stars, rising to 20\% at the innermost regions where $r < 0.1\; h^{-1}$Mpc. 
In fact, the mean pairwise velocity for all the \bahamas simulations, seems to agree at every mass scale regardless of the change in halo mass across each simulation; since this is not a matched set of haloes, the mass binning means that a slightly different set of haloes are selected from each simulation. 
With matched haloes, the differences between the simulations would be reduced even further, but since the differences are quite small to begin with, we have chosen not to show these curves in order to reduce the clutter on this Figure.

Figure~\ref{fig:mean_vr} shows that haloes with feedback have become less compact in the inner regions; that is their mean radial velocities are overall lower within the halo radius when considering the total matter content. 
This is consistent for all the halo mass bins at all radii, except for the very inner region of the least massive haloes that we consider in the 10$^{12}  < h^{-1}$ M$_{\odot} \leq  10^{12.50}$ range, where there is an excess at $r < 0.2 \,h^{-1}$Mpc due to adiabatic contraction, as these haloes have the greatest star formation efficiency but weak AGN feedback. 
Particles within the interior of the halo tend to have lower infall velocities in the \bahamas AGN simulations than in the DM only simulation, with a minimum around the radius where the multistream region starts to appear; for the lower mass bins, this is substantially greater than the virial radius while for the larger bins, they are approximately the same. 
This point is pushed to a slightly larger radial distance in the hydrodynamical simulations. 
This is a well documented effect that is consistent with many previous studies (see for example~\citealt{schaller15}) and has been observed for \bahamas in the aforementioned~\citealt{kuruvilla20}). 
This would imply that the redshift space clustering weakens with the inclusion of baryonic physics, and indeed the monopole and quadrupole power spectra are suppressed on scales $k > 1\,h$ Mpc$^{-1}$ to a slightly lesser degree than in real space, as seen in Figure~\ref{fig:eagle_comparison}.

Across mass bins, however, Figure~\ref{fig:mean_vr} shows that the total mean infall velocity in the haloes with full gas and stellar physics approaches the behaviour of the dark matter only case as the mass of the host increases. 
For very massive haloes, e.g. those in the 10$^{13.50}  <  h^{-1}$ M$_{\odot} \leq 10^{14.00}$ range, we can see that the halo velocity profile deviates the least (with some noise) from its dark matter only counterpart. 
For these haloes, the baryon fraction contained within the cluster approaches the universal value.
This would imply that the redshift space clustering of the largest haloes would be the least impacted by baryonic effects, except we have already seen that the difference in the mean bulk halo peculiar velocity is the greatest for the heaviest haloes.
Related to the discussion of the multistream region, there is also a point where the ratio reaches a minimum value for both the gas and total $\left < v_{12} \right >$ and then increases again. 
The gas infall velocity profile more or less traces that of the total matter content near the halo radius and on larger scales, except there is a peak in the inner regions. 
This feature is absent in the measurements of the stellar $\left < v_{12} \right>$ but there is a peak that occurs at a smaller radius. 

The changes in $\left < v_r \right>$ due to baryonic effects and subgrid physics have already been described by~\citet{kuruvilla20} for all particle tracers in the \bahamas simulations. 
\citet{kuruvilla20} showed that the particle velocity profile is suppressed on all scales against gravity only simulations by only 2-3\% in \bahamas on intermediate scales of 1 $ < r < 10 h^{-1}$Mpc (and no discernible suppression in \eagle) and rising to 7\% in the $r < 0.5 \;h^{-1}$Mpc range (20\% in \eagle).~\footnote{Numbers quoted are for all particles, the gas particles typically show a much greater suppression, about 30\% at its maximum, see Fig 2 (top panel) of~\citet{kuruvilla20}} 
Furthermore, a non-zero velocity bias was observed; for even the greatest separations of $r > 10 \; h^{-1}$Mpc, the ratio between the mean pairwise velocity for the fiducial AGN and DM only simulations did not approach unity. 

Figure~\ref{fig:mean_vr} shows that while our measurements for the total $\left < v_{12} \right>$ for most halo masses are qualitatively similar to~\citet{kuruvilla20}, quantitatively the effects of baryons on the mean halo velocity profile are much stronger. 
For example, even on intermediate scales of $r \sim 0.5 \;h^{-1}$Mpc, we find that $\left < v_{12} \right>$ measured from all halo particle types in the fiducial AGN simulation relative to the dark matter only case are suppressed by as much as 15\% for the two lowest mass bins, which rises sharply to a $>20$\% excess at $r< 0.01\,h^{-1}$Mpc for the haloes in the 10$^{12}  <  h^{-1}$ M$_{\odot} \leq 10^{12.50}$ mass range. 
However, these variations are gradually reduced as the halo mass increases and the baryon fraction captured within the halo approaches the universal value. 
For the gas, we also see a sharp peak in the pairwise velocities after an initial suppression, for example, for the 10$^{12}  <  h^{-1}$ M$_{\odot} \leq 10^{12.50}$ mass bin this is around $r\sim 0.2 \,h^{-1}$Mpc. 
Some of these differences can be attributed to the fact that~\citet{kuruvilla20} uses all the matter particles for their calculation of the mean pairwise velocity statistic, whereas we only consider pairs between halo centers and their associated halo particles. 
Using the full particle distribution tends to wash out features within the halo as pairs between free particles (not bound to any halo) will have been included as well as their cross terms. 
However, the mean halo infall velocity is arguably more relevant for studies that are sensitive to the motion of the gas distribution in clusters such as the kinetic Sunyaev-Zeldovich effect~\citep{sz72,sz80}. 
Furthermore, outside of the halo radius, we draw a different conclusion to~\citet{kuruvilla20}: it is not immediately apparent from our comparison that there is a large scale velocity bias between the haloes formed in the \bahamas simulations with full baryon physics and the dark matter case. 
This is in contrast to~\citet{kuruvilla20} in which the particle mean pairwise velocities are consistently below their dark matter equivalents up to about 1\% on large scales for all particles and 4-5\% when considering the gas only.
However, it is important to note that our halo velocity profiles are only relevant for the 1-halo term and the 2-halo term (more aptly measured by particle-particle pairs) is expected to dominate on large scales. 
But as we saw in the previous section, it is the difference in the halo mass function between the fiducial AGN and the gravity only \bahamas simulations that drive the changes to the redshift space power spectrum on large-scales and these differences are percent level for a matched halo catalogue. 

\begin{figure*}
    \centering
    \includegraphics[width=0.8\linewidth]{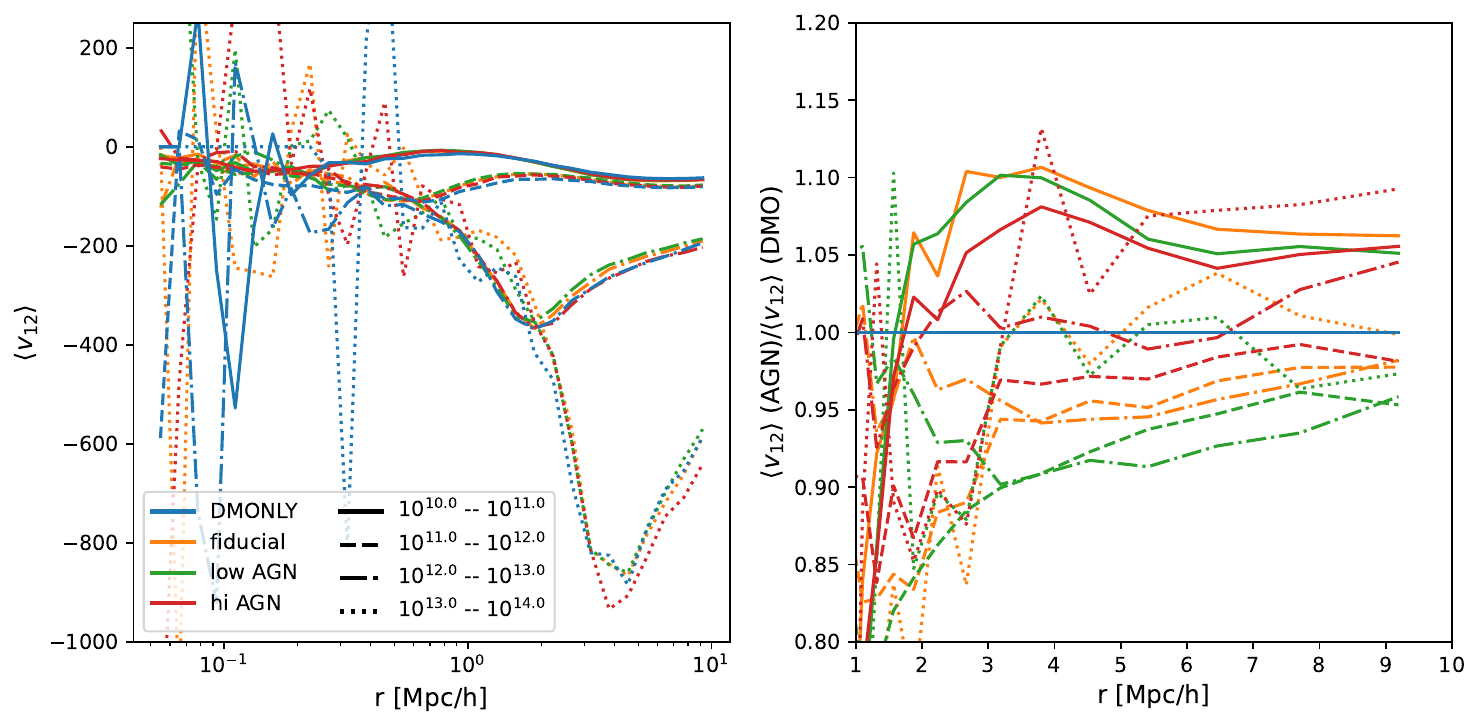}
    \caption{On the left, the mean radial infall velocity calculated between central (group haloes) and their satellites for all four \bahamas simulations for stellar mass bins of 10$^{10}$ < M$_{\star}$ M$_{\odot}\, h^{-1}$ < 10$^{11}$ (solid), 10$^{11}$ < M$_{\star}$ M$_{\odot}\, h^{-1}$ < 10$^{12}$ (dashed), 10$^{12}$ < M$_{\star}$ M$_{\odot}\, h^{-1}$ < 10$^{13}$ (dot-dashed), and 10$^{13}$ < M$_{\star}$ M$_{\odot}\, h^{-1}$ < 10$^{14}$ (dashed). On the right is the ratio of the three AGN \bahamas simulations with respect to the DM only case for each of these stellar mass cuts. Due to the lack of subhaloes occupying the inner regions of the central halo, the measurement is very noisy at scales of $r < 1$ Mpc $h^{-1}$ and so we have only shown the halo outskirts on the right.}
    \label{fig:mean_vr_stellar_mass}
\end{figure*}

In Figure~\ref{fig:mean_vr_stellar_mass}, we measure the mean radial infall velocity for subhaloes binned according to stellar mass. 
Using the same matched catalogues as previously considered for the analysis of the halo-halo power spectrum, we are able to split the halo sample from the four \bahamas simulations into four stellar mass bins logarithmically spaced from 10$^{10}$ M$_{\odot}\, h^{-1}$ (the limit of our stellar mass resolution) to 10$^{14}$ M$_{\odot}\, h^{-1}$ with the highest stellar mass halo at 10$^{13.6}$ M$_{\odot}\, h^{-1}$. 
Because each tracer is now a subhalo of the main central halo, the measurement is much noisier than that of Figure~\ref{fig:mean_vr} which used the full halo particle distribution. 
Moreover, the difficulty of subhalo identification and survival near the center of the group potential means that the inner regions cannot be well sampled.
Nonetheless we can still draw some conclusions about the mean infall velocity near the outskirts of the halo and beyond.
The difference between each AGN simulation and the DM only case has been enhanced by selecting on their stellar content and at its maximum at $r\sim3\,h^{-1}$Mpc, there is a boost of 7--10\% across all models for 10$^{10}$ -- 10$^{11}$  M$_{\odot}\, h^{-1}$ stellar mass haloes and a suppression $\sim$5\% for 10$^{11}$ -- 10$^{12}$  M$_{\odot}\, h^{-1}$ stellar mass haloes. 
We note that observable galaxy samples are likely to come from this range.  
Furthermore, this figure suggests the existence of a large scale velocity bias as the ratios of the mean infall velocity does not seem to converge to zero on scales approaching 10 Mpc $h^{-1}$ contrary to what was measured using the halo particle distribution only.
This halo velocity bias is the strongest for the least massive haloes (in stellar mass) which show a halo bias of $\sim$5\% at the largest scales. 
This is stronger than the percent-level bias observed by~\citet{kuruvilla20} obtained from using all the particles in the simulation. 
There seems to be no discernible trend with the strength of the AGN feedback, however, the low AGN model mostly shows the most extreme behaviour among the set.

\subsubsection{The 1-halo term: Halo density profile}\label{sec:xi1h}
Except through direct measurements of the local velocity field, the infall velocity profile is not directly observable and must be probed via clustering statistics observed in redshift space. 
In this section, we explore how the differences in the mean infall velocities might translate into measurements of the 1-halo correlation function as defined between central-satellite pairs. 
\begin{figure*}
    \centering
    \includegraphics[width=0.8\linewidth]{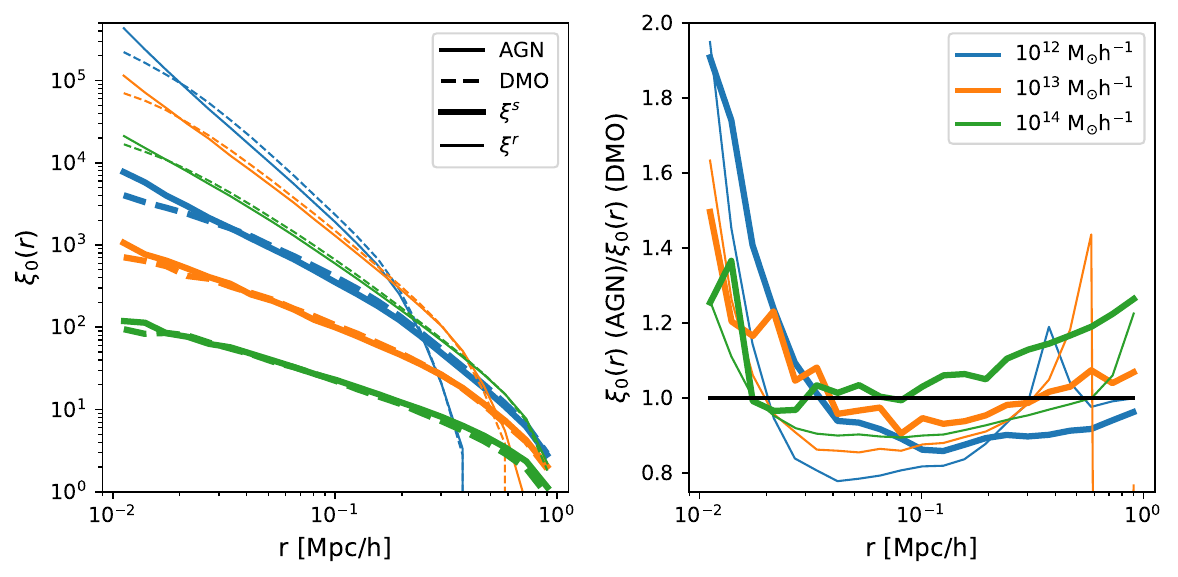}
    \caption{The central-satellite 1-halo correlation function in real (thin lines) and redshift space (thick lines) from \bahamas for 10$^{12}$, 10$^{13}$ and 10$^{14}\,h^{-1}\;$M$_\odot$ haloes. The left panel shows the AGN (solid) and dark matter only (dashed) correlation functions for all particle species as a function of halo mass. The right panels shows the ratio with respect to the dark matter only simulation for each mass bin. Notice the absence of a suppression in the halo profile in redshift space clustering for the highest mass haloes.}
    \label{fig:xi1h}
\end{figure*}

We take the correlation function as pairs between the center of the halo and the particles identified as belonging to the halo as in the definition of the mean infall velocity and use the Landy-Szalay estimator~\citep{landyszalay93} as follows:  
\begin{equation}
\xi_{1h}({\Delta r}) = \frac{DD({\Delta r})- 2DR({ \Delta r}) + RR({\Delta r})}{RR({ \Delta r})}, 
\end{equation}
where $DD$, $DR$, and $RR$, represent the counts of data-data, data-random, and random-random pairs within a bin of width $\Delta r$. 
For the random points, we have chosen a uniform distribution across $\pm 2 \; h^{-1}$ Mpc with at least 20 times the total number of halo particles in each sample. 
For Figure~\ref{fig:xi1h}, we only consider haloes of mass 10$^{12}$, 10$^{13}$ and 10$^{14} h^{-1}\;$M$_\odot$, where the bin width has been chosen such that there are at least 500 members in each, and stacked these in order to improve the signal-to-noise of the measurement of the correlation function. 
This necessitates the use of the Landy-Szalay estimator since we can no longer assume periodic boundary conditions around the stacked halo. 
For the redshift space measurements, we use Equation~\ref{eqn:particleshift} to convert the positions of halo centers and their particles to redshift space. 
The 1-halo correlation functions have been measured using the \texttt{TreeCorr} package~\citep{jarvis04} for speed. 

Figure~\ref{fig:xi1h} shows the 1-halo central-satellite correlation function in both real and redshift space as measured from the fiducial AGN and dark matter only \bahamas simulations. 
There are a number of interesting points about this figure; firstly the behaviour of the 1-halo term in the fiducial AGN simulation is quite different between real and redshift space. 
The fiducial AGN real space 1-halo correlation function shows the typical suppression of the halo profile that has been pointed out by many authors previously e.g.~\citet{schaller15}, namely that material has been `scooped out' by AGN feedback from the intermediate $r \sim 0.1\;h^{-1}$Mpc region of the halo.
However, this feature is much smaller in redshift space, in which the suppression is roughly half the size as in real space and gradually reduces with increasing halo mass; in fact, for the 10$^{14} h^{-1}\;$M$_\odot$ haloes there is no longer a suppression but an enhancement instead. 
This is consistent with the overall effect of baryons on the redshift space matter power spectrum; in Figure~\ref{fig:eagle_comparison}, we can see that the suppression of the monopole and quadrupole moments on scales $k > 1 \, h^{-1}$Mpc is weaker than the suppression on the real space power spectrum. 
Also, given the behaviour of the infall velocities shown in Figure~\ref{fig:mean_vr}, we would expect the 10$^{14}h^{-1}\;$M$_\odot$ haloes to show the least difference between the AGN and dark matter only simulations, however, the 1-halo correlation function is a mass weighted quantity and so gives greater prominence to the dark matter particles.
From Figure~\ref{fig:xi1h}, we can also see that redshift space distortions have smeared out the clustering near the center of the halo, due to motions of the stellar component as well the multistreaming region near the `edge' of the halo. But the enhancement of clustering due to the internal velocities near the centre of the halo ($r < 0.3 \,h^{-1}$Mpc) is much more extreme in the 1-halo correlation function than the mean infall velocity across all mass ranges.

\begin{figure*}
    \centering
    \includegraphics[width=0.8\linewidth]{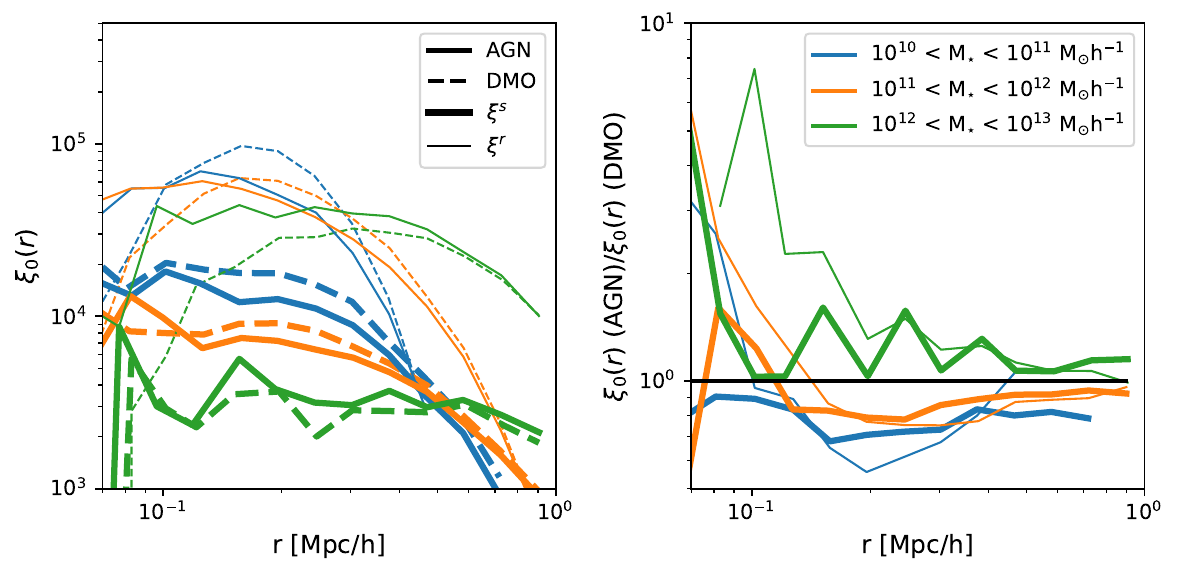}
    \caption{The central-satellite 1-halo correlation function in real (thin lines) and redshift space (thick lines) from \bahamas for haloes of 10$^{10}$, 10$^{11}$ and 10$^{12} \, h^{-1}\;$M$_\odot$ in stellar mass. As in Figure~\ref{fig:xi1h}, the left panel shows the AGN (solid) and dark matter only (dashed) correlation functions and the right panel shows the ratio with respect to the dark matter only simulation for each mass bin.}
    \label{fig:xi1h_stellar_mass_selected}
\end{figure*}

In Figure~\ref{fig:xi1h_stellar_mass_selected}, we have binned the central-satellite pairs into the same four stellar mass bins as previously, starting from 10$^{10}$ to 10$^{13}$ M$_{\odot}\,h^{-1}$ in stellar mass. 
Each satellite is now a subhalo and we take the central as the centre of the FOF group identified by \textsc{Subfind}.
Again we stack on the group position which is taken to be at rest with respect to the satellite members belonging to the group, that is the redshift space distortions in this section are calculated from the difference between the satellite and group velocities in order to isolate the 1-halo term. 
For readability, bins in $r$ have been excluded from Figure~\ref{fig:xi1h_stellar_mass_selected} where there were too few counts near the edge of the halo (for less massive centrals) and near the center of potential for more massive haloes, where satellites are more likely to be destroyed. 
While there were haloes with stellar masses greater than $10^{13}$ M$_{\odot}\,h^{-1}$, these were not sufficient in number to obtain a correlation function that held any meaningful results. 
As in the previous case with mean infall velocities, we find that taking central-satellite pairs instead of central-particle pairs rather enhances the effect of baryons on the observable in question.
Note the change in the scale in the right panel of Figure~\ref{fig:xi1h_stellar_mass_selected} from Figure~\ref{fig:xi1h}, the amplitude of the 1-halo correlation function in the regions close to the center are now several factors of the DM only result.
The satellite haloes themselves are located in density maxima and so their peculiar velocities are local extrema, which serves to make differences in observables measured in redshift space appear stronger.
As for the central-satellite particle pairs, the \bahamas AGN tuned redshift space 1-halo correlation function is suppressed relative to the same dark matter measurement. There is a $\sim$20\% suppression for 10$^{10}$ -- 10$^{12} \, h^{-1}\;$M$_\odot$ stellar mass haloes relative to the DM only case for $r > 0.1 \, h^{-1}$Mpc while the real space results look qualitative similar to those in Figure~\ref{fig:xi1h}; however there is a stronger enhancement at small radii leading to a suppression around $0.2 < r < 0.4 \, h^{-1}$Mpc for the lower stellar mass bins. 

As the stellar mass of the sample increases, Figure~\ref{fig:xi1h_stellar_mass_selected} shows that the difference between the dark matter only and AGN simulations in redshift space decreases, which is consistent with our earlier findings involving the 2-halo monopole power spectrum. 
These differences could pose as a potential systematic when measuring the 1-halo correlation function in a galaxy sample. 
However, a more quantitative analysis would require the use of a larger sample size of haloes and in turn large simulations, which we defer to future work.

\subsection{Volume effects}\label{sec:boxsize_test}
We explore the effect of the size of the simulation volume on the baryonic impact on the redshift space power spectrum. 
This is important to address because of the volume difference between \bahamas and \eagle, as larger simulation volumes are more likely to contain cluster-sized haloes, because these are rare objects. 
This has a impact on the power spectrum since it is these haloes that make the most significant contributions to clustering on scales $2 < k < 10 \, h\,$Mpc$^{-1}$~\citep{vandaalen15}. 
Furthermore, we have shown that the peak in the ratio of quadrupole moments is related to the transition regime between Kaiser squashing and the non-linear Fingers-of-God effect. 
If the simulation box size is insufficient to capture this feature, then is expected that much of the signal in the quadrupole would be removed. 

For these tests, we use the \textsc{Antilles} simulations with a similar feedback model and mass resolution as the \bahamas fiducial case but at three different box sizes, namely 100, 200, and 800 $h^{-1}$Mpc a side.
Although these runs that were part of a calibration campaign for~\citet{salcido23}, their subgrid parameters are similar enough to \bahamas such that the main features between the two sets are nearly indistinguishable in Figure~\ref{fig:boxsize_test} across all scales. 

We present the residuals on the monopole and quadrupole moments between their dark matter only counterparts as a function of varying the simulation volume in Figure~\ref{fig:boxsize_test}. 
The largest volume, \textsc{Antilles}-L800, shows that the effects of gas physics only dips below percent level for $k \sim 0.1 \,h\,$Mpc$^{-1}$ in the monopole and $k \sim 0.07 \,h\,$Mpc$^{-1}$ in the quadrupole. 
From this figure, we note that the monopole moment is similar between all the simulations regardless of their size and that the peak at $k\sim0.15\,h\,$Mpc$^{-1}$ in the quadrupole is a common feature to all the larger volumes but not \eagle which has a box length of 100 Mpc (or 67.77 $h^{-1}$ Mpc) per side. 
This feature in the quadrupole is noticeably smaller for the \textsc{Antilles}-L200 and {\textsc{Antilles}-L100 boxes. 
From this, we can conclude that volumes below (200 $h^{-1}$Mpc)$^3$ are insufficient to capture the full impact of baryonic physics on the redshift space matter power spectrum. 
Notice that the \textsc{Antilles}-L100 simulation fails to capture the full extent of the transition between linear and non-linear regime in the quadrupole and the magnitude of the residuals are dampened compared to the larger box sizes. 

However, while we have shown that the volume of the simulation significantly affects the measurement of baryonic effects on the quadrupole moment, this cannot completely account for the difference between \eagle and \bahamas, since at $k>1\,h\,$Mpc$^{-1}$, where the box size should no longer have an effect, we still observe a substantial difference between the two simulations. 

Furthermore, the monopole moments are well converged over the full range of volumes that we have tested, and yet the monopole moment in \eagle is still suppressed relative to \bahamas. Both of these effects are likely driven by the differences in the feedback calibration strategy as already discussed.

\begin{figure}
    \centering
    \includegraphics[width=\linewidth]{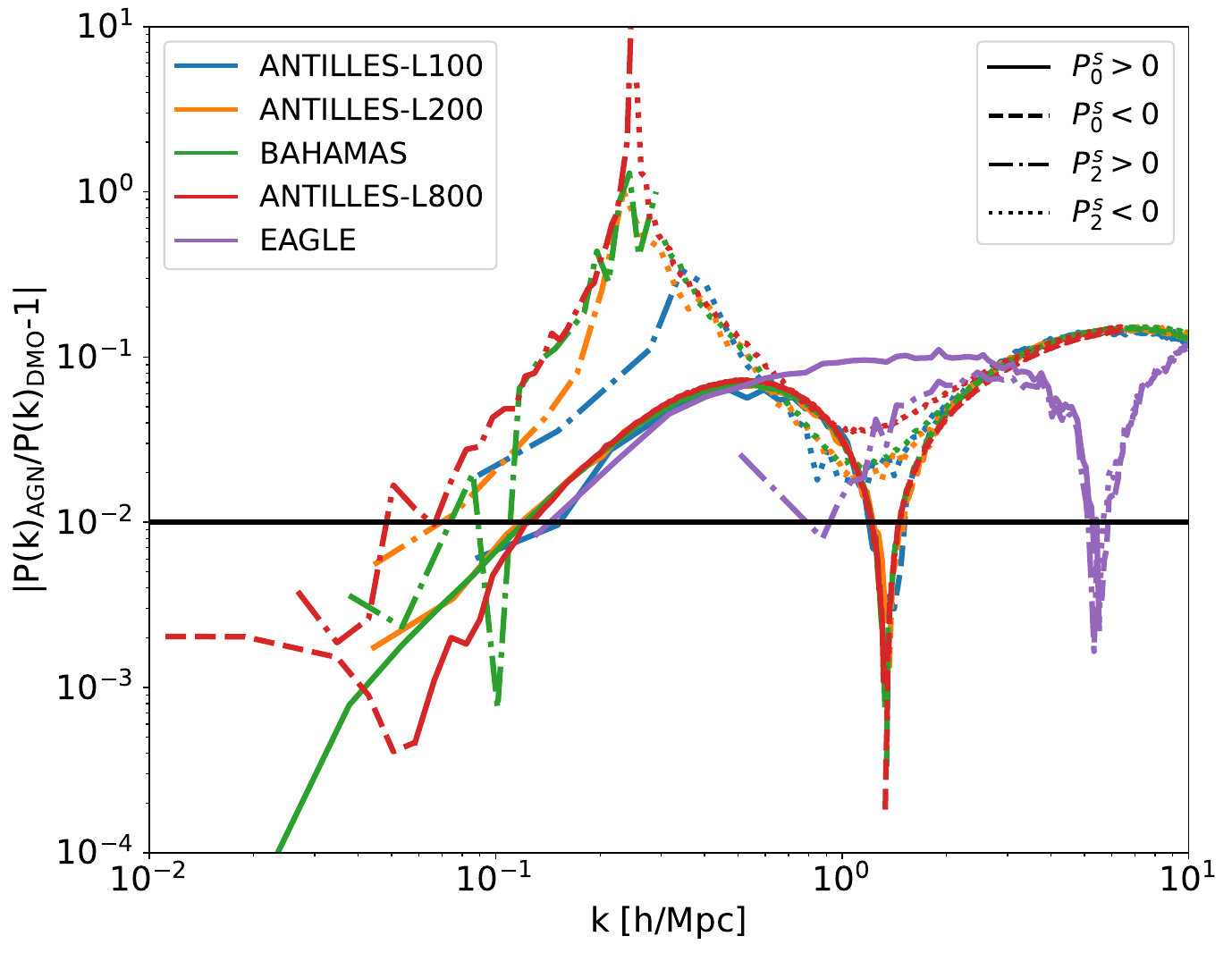}
    \caption{A test of the effect of the simulation box length (in $h^{-1}$ Mpc) on the measured residuals between the dark matter only and AGN reference monopole and quadrupole power spectra using the calibrated \textsc{Antilles} simulation boxes. We also show the \eagle results (L = 67.77 $h^{-1}$ Mpc) for comparison.
    }
    \label{fig:boxsize_test}
\end{figure}

\subsection{Redshift dependence}\label{sec:z_dependence}

Our previous results have been obtained from a comparison at $z=0$, but we now investigate the measurements of the monopole and quadrupole from two additional snapshots, namely $z=0.5$ and $z=0.75$, that are relevant for cosmologically interesting galaxy samples, such as the CMASS selection~\citep{reid16} at $z=0.5$, and also eBOSS Emission Line Galaxies (ELG) galaxies at higher redshifts~\citep{delubac17}.} 
Figure~\ref{fig:rsd_z_dependence} shows the change in the residuals between the fiducial AGN and DM only simulations in the monopole and quadrupole moments as a function of redshift. 
The difference in the monopole is consistently around a few percent on the quasi-linear scales, peaking at $\approx 10$\% for $k\sim 0.5 \,h\,$Mpc$^{-1}$ for all redshifts considered. 
Meanwhile the spike in the residuals of the quadrupole moment is present at every redshift, and moreover moves to smaller scales as the redshift increases; this is consistent with its origin being a change in sign from the onset of non-linear redshift space distortions.
On large scales, the difference in the quadrupole moments appears less important, dropping to 2\% with $z=0.75$ at $k \sim 0.1 \,h\,$ Mpc$^{-1}$. 
However, Figure~\ref{fig:rsd_z_dependence} does show that the quadrupole is consistently more sensitive to baryonic physics than the monopole on the quasi-linear scales of $ 0.1 < k < 1 \, h$ Mpc$^{-1}$, while both are similarly affected deep in the non-linear regime of clustering at $k \sim 10 \, h$ Mpc$^{-1}$. 
Figure~\ref{fig:rsd_z_dependence} suggests that the baryonic effects on the quadrupole moment on large-scales do not really become below sub-percent level (and thus can be safely ignored) until much higher redshifts than $z = 0.75$ are considered.

\begin{figure}
\includegraphics[width=\linewidth]{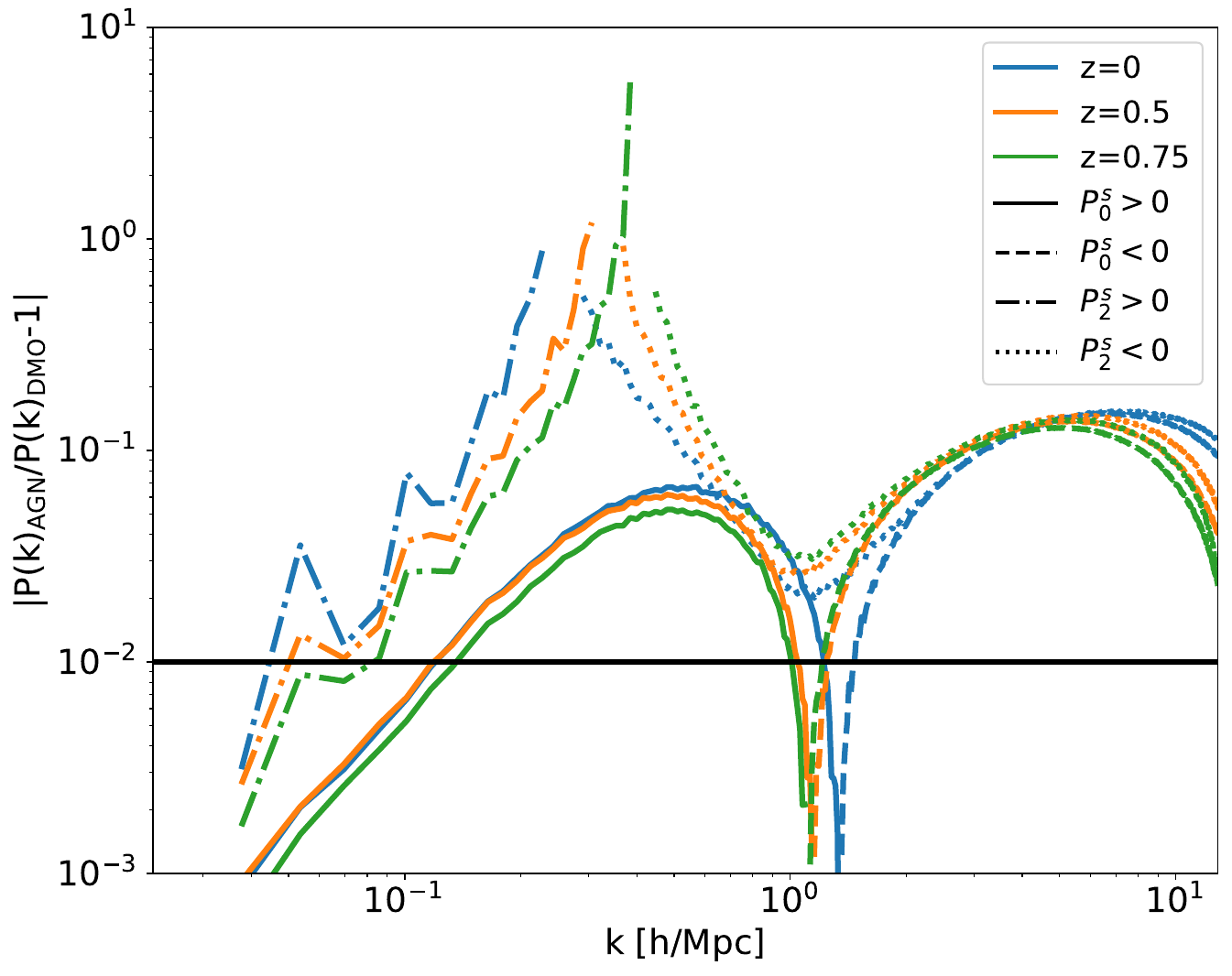}
\caption{The ratio between monopole and quadrupole power spectra for the fiducial AGN and DM only simulations measured at three different redshifts: $z=0$ (blue), $z=0.5$ (yellow) and $z=0.75$ (green). Since we have shown the absolute value of the difference, solid (dot-dashed) lines represent the positive values of the ratio for the monopole (quadrupole) moment, while dashed (dotted) lines represent negative values of the ratio for the monopole (quadrupole) moment.}
\label{fig:rsd_z_dependence}
\end{figure}

\subsection{Dependence on baryon modeling}\label{sec:feedback}
We now explore in greater depth the effect of changes in modelling baryonic physics to the redshift space monopole and quadrupole by comparing our measurements from \bahamas to the \textsc{cosmo-OWLs} simulation suite. 
As mentioned in the introduction, the purpose of \textsc{cosmo-OWLs} was to run a variety of models with various gas physics turned on and off to see what the effect would be on large-scale structure observables. 
The \bahamas suite also includes an additional two simulations that bracket the calibrated value of the AGN heating temperature, $\log_{10}\Delta$T = 7.8: $\log_{10}\Delta$T = 7.6 (`low AGN') and $\log_{10}\Delta$T = 8.0 (`hi AGN') and hence the observed gas fractions of galaxy groups, while \textsc{cosmo-OWLS} contains more extreme variations, such as simulations without any radiative cooling (NOCOOL), no AGN feedback (REF) and several different AGN heating temperatures, $\log_{10}\Delta$T = 8.0, 8.5 and 8.7 (referred to as AGN 8.0, AGN 8.5, and AGN 8.7 hereafter). 
There is much more variation in the gas physics encompassed by the \textsc{cosmo-OWLs} simulations in this section than in the comparison between \bahamas and \textsc{Eagle} in Figure~\ref{fig:eagle_comparison}, with the lack of radiative cooling, star formation, and feedback being the most extreme cases. 
In Section~\ref{sec:matterpk}, we have seen that one of the major differences between the \bahamas and \eagle simulations is their approach to modelling subgrid physics. 
The \bahamas suite is calibrated to both the local galaxy stellar mass function and the gas fraction in clusters, while \eagle is only calibrated to the galaxy stellar mass function but predicts group gas fractions that are larger than observed.

\begin{figure*}
    \centering
    \includegraphics[width=\linewidth]{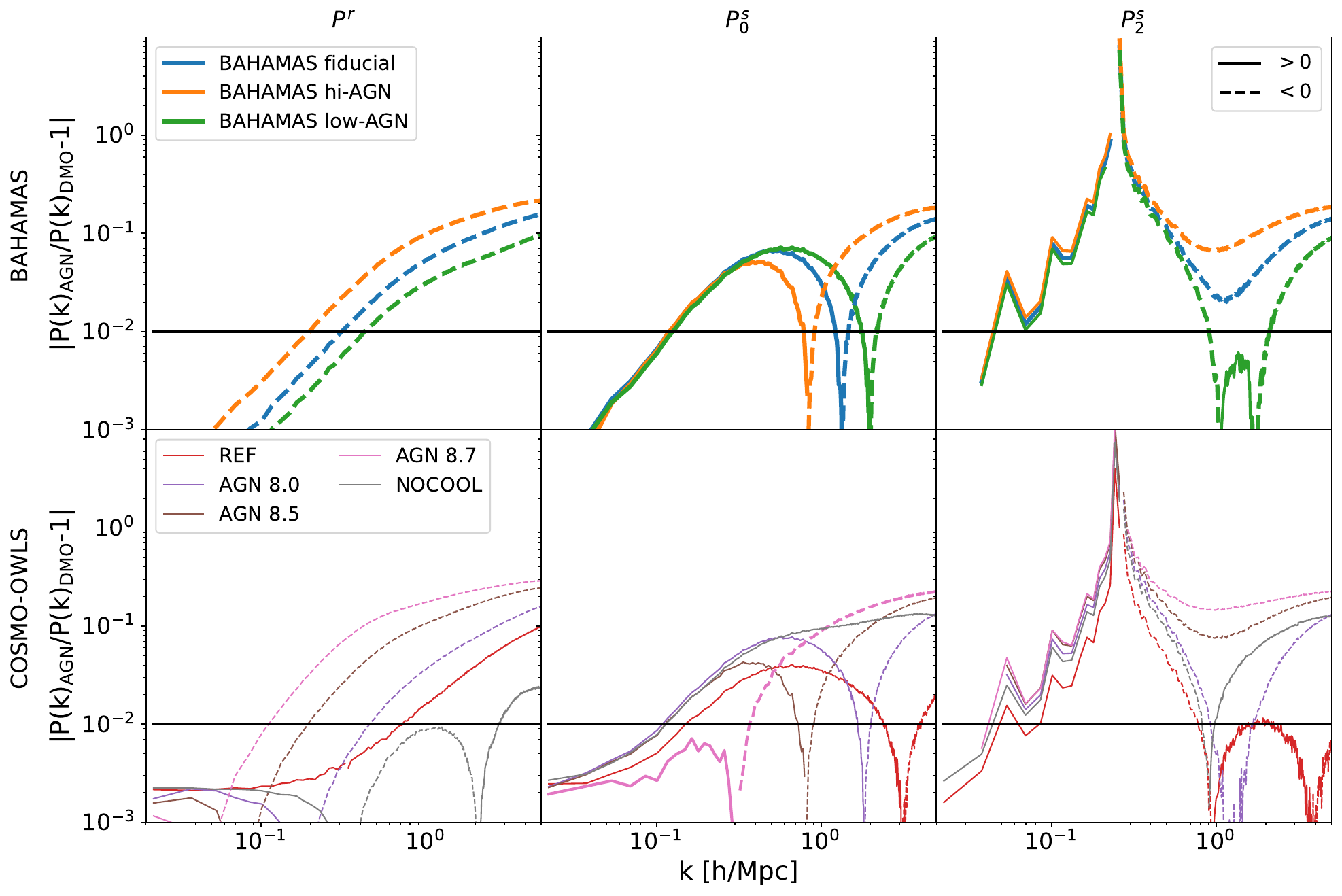}
    \caption{Effect of varying subgrid physics on the real space (left) monopole (middle) and quadrupole (right) power spectra as encapsulated by \bahamas (top row, thick lines) and \textsc{cosmo-OWLS} (bottom row, thin lines). Solid lines denote positive values, while dotted lines are used for negative values of the residual. Note that hi-AGN and low-AGN \bahamas models only have a single realisation and in order to match this, we have opted to only show the same realisation for the AGN fiducial simulation as well.}
    \label{fig:cosmoowls}
\end{figure*}

Figure~\ref{fig:cosmoowls} shows the ratio of the real space, monopole and quadrupole power spectra in three panels from left to right for each simulation that we consider in this section. 
The fiducial \bahamas value of $\log_{10}\Delta$T = 7.8 are shown as blue curves, while the two extremes $\log_{10}\Delta$T = 7.6 and $\log_{10}\Delta$T = 8.0 are shown as yellow and green lines respectively in Figure~\ref{fig:cosmoowls}. 
The \textsc{cosmo-OWLs} suite supplies another three AGN heating temperatures, $\log_{10}\Delta$T = 8.0 (purple), $\log_{10}\Delta$T = 8.5 (brown) and $\log_{10}\Delta$T = 8.7 (pink), but it is difficult to draw conclusions based on the values of $\log_{10}\Delta$T alone across the two simulation suites because other subgrid parameters have been varied as well. 
Nonetheless, comparing the behaviour of these two-point functions shows that the redshift space monopole and quadrupole are less sensitive to the individual details of the baryon physics in each simulation (such as the value of $\log_{10}\Delta$T) than in real space. In Figure~\ref{fig:cosmoowls}, changing the AGN heating temperature seems to shift the real space power spectrum by an almost constant magnitude across all scales, whereas the monopole and quadrupole moments seem to show less variation overall. In fact most of these models agree on large scales and only appreciably differ above $k\sim 0.2\,h\,$Mpc$^{-1}$, where the 1-halo term starts to become significant. 
For some \textsc{cosmo-OWLs} simulations, e.g. REF, the convergence on large scales is slower, because the subgrid modelling has been varied by a greater amount, in this case, the REF simulation doesn't involve any AGN feedback.
In \bahamas, the presence of any baryonic effects at all boosts the large scale ($k < 0.2\,h\,$Mpc$^{-1}$) redshift space multipoles relative to the dark matter only case but by a similar amount independent of the AGN temperature and hence the amount of material being ejected. 
Furthermore, the peak in the quadrupole residuals appears at approximately the same $k$ regardless of any of the extreme feedback scenarios considered within the \textsc{cosmo-owls} suite.  
This is consistent with our conclusion that the difference in feedback schemes between \bahamas and \textsc{Eagle} are not primarily responsible for this feature. 
While the 2-halo redshift space clustering is less sensitive to the details of AGN feedback, the effect of baryons start to take effect at a lower $k$ value than in real space - the leftmost plot shows that the power spectrum from the fiducial AGN simulation reaches a percent level deviation from the dark matter only simulation on scales below $k \sim 0.3\,h\,$Mpc$^{-1}$ in real space but the corresponding scale for the redshift space monopole moment is at $k \sim 0.1\,h\,$Mpc$^{-1}$.

Having shown that varying the prescriptions for feedback has a smaller impact on the redshift space than on the real space power spectra in the linear and quasi-linear regimes, we now examine their effects for $k>0.2\,h\,$Mpc, where most simulations start to disagree.
In fact, we find that on small scales, the power spectra across the different feedback prescriptions are also less impacted in redshift space than in real space on $k$ scales that probe well into the halo profile.
Taking the ratios of each power spectra in Figure~\ref{fig:cosmoowls} with respect to the \bahamas fiducial simulation reveals that at most there is a 15.5\% variation in the monopole that occurs at $k=1.6\,h\,$ Mpc$^{-1}$. For the quadrupole, this is an 18\% variation at $k=1.2\,h\,$Mpc$^{-1}$, where we have excluded the $k$ scale where the quadrupole changes sign.
For comparison, in real space, the maximum difference is 23.7\% at $k=4.6\,h\,$Mpc$^{-1}$. 
These figures are obtained from the \textsc{cosmo-OWLS} AGN 8.7 simulation, which we can see from Figure~\ref{fig:cosmoowls} is the most discrepant model across all three summary statistics.

\section{Conclusions}
\label{sec:conclusions}
Our investigation of the behaviour of the monopole and quadrupole moments in redshift space in the \bahamas suite of simulations have allowed us to quantify the impact of baryonic effects on the redshift space matter and halo two-point clustering statistics. 
We have compared the \bahamas redshift space power spectra to those measured from the \eagle simulations, investigating the 2-halo term as a function of group mass and the 1-halo term by characterizing both the mean infall velocity of particles inside groups and the distribution of centre-satellite pairs as a function of particle species and stellar mass. 
Furthermore, we also extend our analysis to different simulation volumes with the \textsc{Antilles} suite and extreme feedback scenarios with the \textsc{cosmo owls} suite as well as two additional \bahamas simulations with stronger and weaker AGN feedback.

Our examination of the redshift space clustering in the \bahamas suite has led us to the following conclusions:
\begin{itemize}
    \item We found that contrary to~\citet{hellwing16}, baryonic effects can have a greater than percent level impact on the redshift space matter power spectrum on scales as large as $k \sim 0.1\,h$Mpc$^{-1}$. This is because the \eagle simulation fails to both capture the true extent of gas physics and the transition region between the linear Kaiser model to the non-linear `Fingers-of-God' effect. Furthermore, this effect is scale-dependent and introduces an additional bias between the total matter and galaxy power spectra, as shown in Figure~\ref{fig:halo_mass_selected}, although the quantitative impact on cosmological parameter estimation has yet to be determined. 
    \item On large scales (around $k < 0.1\; h$Mpc$^{-1}$), changes to the halo mass function from the inclusion of hydrodynamics induces a 7-10\% (3\%) level change to the halo monopole (quadrupole) for haloes in the observationally interesting mass range of 10$^{13} < h^{-1}$M$_\odot \le $10$^{13.5}$. When using a matched set of haloes to calculate the halo monopole and quadrupole moments, then the difference with and without baryonic physics becomes sub-percent level up to $k < 0.3 \,h^{-1}$ Mpc , except when low stellar mass haloes in the range 10$^{10} < h^{-1}$M$_\star \le $10$^{11}$ are considered.
    \item The presence of baryons can alter several features of the mean infall velocity of particles, such as the location of the multistream region as well as particles slowing down in the interior of the halo as the halo concentration is reduced when gas physics is added. However, these features are only marginally affected by changes in feedback model as such as the value of AGN heating temperature, $\log_{10}\Delta$T as seen in the top left panel of Figure~\ref{fig:mean_vr}. When we consider the infall velocities of subhaloes selected on the basis of stellar mass, we observe a 1-5\% velocity bias on scales of $\sim$10 Mpc $h^{-1}$ consistently across all stellar mass bins considered.
    \item The redshift space 1-halo correlation function can show a significant displacement of the distribution of halo particles and subhaloes around the central when the effect of baryons are included. We found a $\sim$20\% suppression at scales $r > 0.2\,h^{-1}$Mpc relative to the dark matter only result for stellar masses in the range 10$^{10}$--10$^{12}$ M$_\odot\,h^{-1}$. 
    \item Figure~\ref{fig:boxsize_test} shows that the simulation volume plays an important role in ensuring that the correct baryonic effects can be deduced. Simulations that are too small, e.g. \eagle, will have difficulties in observing the 2-halo term of the monopole and quadrupole moments, in which the redshift space observables experience an enhancement in clustering, as opposed to a suppression in real space. 
    \item The effects of baryons on quasi-linear quadrupole moment remain at the level of 2-4\% up to intermediate redshifts of $z=0.75$, with a smaller impact on the monopole moment. On small scales ($k < 1 \;h$Mpc$^{-1}$) the impact of baryons on both the monopole and quadrupole are just as significant (at the $>$10\% level) for all the redshifts considered. 
    \item Overall, the redshift space multipoles are less sensitive to the modelling of feedback and implementation of gas physics compared with the real space power spectrum. Many features of the redshift space power spectra are also robust to variations in the subgrid modelling, including the transition from the linear Kaiser regime to non-linear smearing in the quadrupole moment.
 
\end{itemize}
While we have shown the effects of baryons on the redshift space matter  and halo power spectra for stellar selected samples, it difficult to extrapolate these effects to the galaxy redshift space two-point since these do not directly trace the total matter power spectrum, but may involve a complication function of the halo bias and stellar selection function as well.  
Future work involving the creation of mock catalogues with large volume hydrodynamical simulation suites, such as \textsc{Flamingo}~\citep{schaye23} and Magneticum~\citep{dolag16}, will be crucial for assessing the potential impact upon current analyses of the redshift space multipoles for cosmological parameter estimation.

\section*{Acknowledgements}
The authors would like to acknowledge helpful discussions with Amol Upadhye. This project has received funding from the European Research Council (ERC) under the European Union’s Horizon 2020 research and innovation programme (grant agreement No 769130). This work used the DiRAC@Durham facility managed by the Institute for Computational Cosmology on behalf of the STFC DiRAC HPC Facility (\url{www.dirac.ac.uk}). The equipment was funded by BEIS capital funding via STFC capital grants ST/K00042X/1, ST/P002293/1, ST/R002371/1 and ST/S002502/1, Durham University and STFC operations grant ST/R000832/1. DiRAC is part of the National e-Infrastructure.  

\section*{Data Availability}

The data supporting the plots within this article, including the raw \bahamas and \textsc{Antilles} simulation data are available on reasonable request to the corresponding author. Note that the data volume may prohibit us from simply placing the raw data on a server. In the meantime, people interested in using the simulations are encouraged to contact the corresponding author.

\bsp	
\label{lastpage}
\end{document}